\long\def\papselect#1#2{#1}
\papselect{
 \documentclass[12pt]{article}
 \setlength{\textwidth}{6.3in}
 \setlength{\textheight}{9in}
 \setlength{\oddsidemargin}{0in}
 \setlength{\topmargin}{-0.3in}
 \usepackage{amsmath,amssymb,amsthm}
}
{
 \documentclass[twocolumn]{article}
 \usepackage{imr}
}
\usepackage{epsfig}
\usepackage{amsmath,amssymb,amsthm}
\newcommand\commentout[1]{\relax}
\newcommand\mytitlet{An algorithm for two-dimensional mesh generation based
on the pinwheel tiling}
\newcommand\mythankst{Supported in part by NSF Grants CMS-0239068 and CCF-0085969}
\newcommand\firstauth{Pritam Ganguly}
\newcommand\firstauthfn{Department of Theoretical and Applied Mechanics,
Thurston Hall, Cornell University, Ithaca, NY 14853, U.S.A., pg45@cornell.edu.}
\newcommand\secondauth{Stephen A.~Vavasis}
\newcommand\secondauthfn{Department of Computer Science, Upson Hall Cornell 
University, Ithaca, NY 14853, U.S.A., vavasis@cs.cornell.edu.}
\newcommand\thirdauth{Katerina D.~Papoulia}
\newcommand\thirdauthfn{School of Civil and Environmental Engineering, 
Hollister Hall, Cornell University, Ithaca, NY 14853, U.S.A., kp58@cornell.edu.}
\newcommand\abstr{
We propose a new two-dimensional meshing algorithm
called PINW able to generate meshes that accurately
approximate the distance
between any two domain points by paths composed only of cell edges.
This technique is based on an extension of
pinwheel tilings proposed by Radin and Conway. We prove that the algorithm
produces triangles of bounded aspect ratio. This kind of mesh would be
useful in cohesive interface
finite element modeling when the crack propagation path
is an outcome of a simulation process.
}
\newcommand\kwds{
Pinwheel tiling, mesh generation, isoperimetric, cohesive interface
finite element}
\newcommand\area{\mathop{\rm area}}
\renewcommand\big{\mathop{\rm big}}
\newcommand\devi{\mathop{\rm dev}}
\newcommand\diam{\mathop{\rm diam}}
\newcommand\dist{\mathop{\rm dist}}
\newcommand\eref[1]{$(\ref{#1})$}
\newcommand\length{\mathop{\rm length}}
\newcommand\M{{\mathcal M}}
\newcommand\minalt{\mathop{\rm minalt}}

\newcommand\Sk{\mathop{\rm Skel}}
\newcommand\sref[1]{Section~$\ref{#1}$}
\newcommand\T{{\mathcal T}}
\newtheorem{theorem}{Theorem}
\newtheorem{lemma}{Lemma}
\newtheorem{corollary}{Corollary}
\papselect{
 \bibliographystyle{plain}
 
\begin{document}
 \title{\mytitlet\thanks{\mythankst}}
 \author{\firstauth\thanks{\firstauthfn}
 \and 
 \secondauth\thanks{\secondauthfn}
 \and 
 \thirdauth\thanks{\thirdauthfn}
 }
 \maketitle
 \begin{abstract}
 \abstr
 \end{abstract}
}
{
 \def\thepage {}
 \bibliographystyle{imr}
 \begin{document}
 \edef\uctitle{\expandafter\uppercase{\mytitlet}}
 \title{\uctitle\thanks{\mythankst}}
 \author{\firstauth$^1$ \and \secondauth$^2$ \and 
  	\thirdauth$^3$}
 \date{
 $^1$\firstauthfn \\
 $^2$\secondauthfn \\
 $^3$\thirdauthfn
 }
 %% \abstract{*} and \keywords{*} must be before \maketitle.
 \abstract{
 \abstr
 }
 \maketitle
}
\section{Introduction}
One of the most widely used techniques to simulate fracture is
cohesive interface finite element modeling. In this kind of model,
the area or volume under consideration is subdivided into bulk
elements, 
which are typically triangles
or quadrilaterals in 2D and tetrahedra or hexahedra in 3D.
Next, interfacial elements, which are edge elements in 2D or surface elements
in 3D, are placed between some or all pairs of adjacent bulk elements.
The cohesive model prescribes a relationship
relating traction to relative displacement on the interfacial
elements. 
There is an abundance of
literature that deals with the nature of this relationship, e.g., see
\cite{Pap-Vav} and the references therein. 
A widely accepted modeling assumption is that the total energy
to create the crack is proportional to its surface area
(or length in 2D).  In fact, the 
critical energy release rate $G_c$ per
unit surface area or length of crack is often a parameter
of the cohesive model.

In a finite element model, the energy release rate is associated
with surface area or length of interfacial elements composing
the crack being modeled.
If the
discrepancy between the ``true'' crack path (i.e., the path the crack
would follow if it were not for the finite element constraint that
the crack path must lie on predetermined interfacial elements)
and the path of the
simulated crack is large for certain paths, then 
nonphysical
preferred crack
directions can exist. 
In other words, the results of the simulation would depend
upon how well the  boundaries of the mesh cells are aligned along the true
crack path. In this paper, we propose a meshing technique that
approximates the true path with the path along mesh boundaries with
high accuracy even though the true path is unknown to the
mesh generation algorithm.  In particular,
the approximation has the property that for any crack path,
the simulated and true
crack path lengths converge to each other upon refining
the mesh, which is a property not possessed by other simpler
families of meshes. We call this algorithm
the PINW mesh generator
because it is based on an extension of the 1:2 pinwheel tiling
described in the next section.

In Section~\ref{experiment} we define ``deviation ratio''
and consider a simple experiment to test
the properties of the 1:2 pinwheel mesh.
The 1:2 pinwheel tiling seems to be too restricted to be useful
for a general-purpose algorithm, so we explain how to generalize
it to arbitrary triangles in Section~\ref{sec:generalization}.
This generalization is the basis for our meshing algorithm PINW.
In Section~\ref{proof} we prove that our generalization still
has the isoperimetric property.
Then in Section~\ref{algo} we describe the algorithm.
The main new ingredient introduced in that section is a procedure
to convert a tiling to a mesh.
The aspect ratio of the resulting mesh is analyzed in
Section~\ref{aspeffect}.

The aspect ratio of the 
mesh is important for the cohesive fracture application
because the bulk elements (e.g., triangles in 2D) are used to
model a continuum mechanical theory such as linear elasticity.  It
is well-known (see, e.g., Theorem 4.4.4 of \cite{BrennerScott},
in which aspect ratio is called ``chunkiness'')
that poorly shaped elements can lead to substantial
errors in the elasticity solution.

\section{Pinwheel tilings}
In this section,
we provide a brief introduction to the properties of
pinwheel tilings. Tilings are a covering of the euclidean 2-space
${\cal E}^2$ starting with a finite number of shapes called
\emph{prototiles}. The tilings are constructed by translated and
rotated copies of the prototiles that intersect each other only along
the boundaries. The tilings were proposed to model
crystallographic structures in the physics community.
 
The pinwheel tilings \cite{Radin94} are classified as
\emph{aperiodic tilings}. In ${\cal{E}}^2$ this is equivalent to
saying that no translation of the tiling leaves it invariant. The
basic pinwheel tiling as developed by Radin and Conway
has a hierarchical structure
and is constructed by successive operations of subdivisions and
expansions.

\begin{figure}[htb]
\begin{center}
\epsfig{file=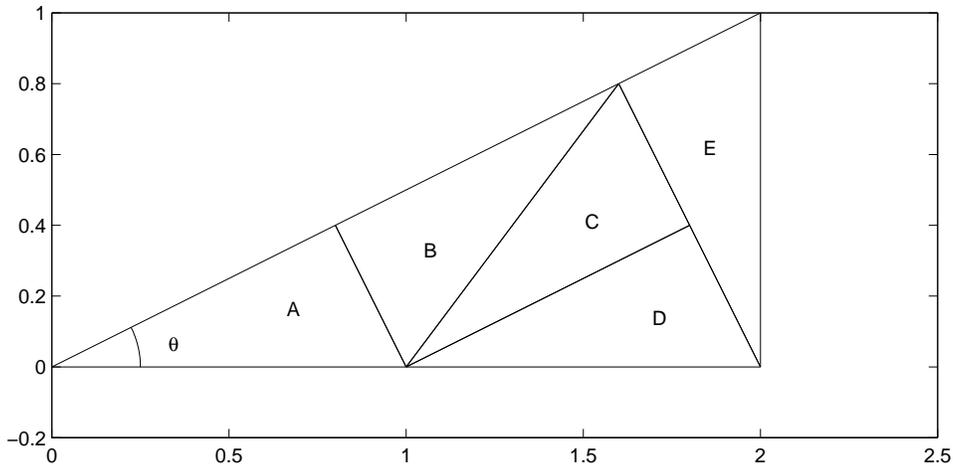, width=0.8\columnwidth}
\end{center}
\caption{Basic pinwheel subdivision proposed by Radin.}
\label{fig:basicPin}
\end{figure}

Consider a right triangle with legs of length $1$ and $2$ referred to
as the \emph{short} and \emph{medium} sides. The hypotenuse is thus of
length $\sqrt{5}$ and will be called the \emph{long} edge. The
vertices will be named similarly, that is, the \emph{small}, \emph{medium}
and \emph{long} vertices are opposite the corresponding sides. For
brevity, we will call a right triangle with the ratio of its short to
medium edge equal to  $1/2$ as a ``$1:2$ right triangle'' and the tiling
formed by its copies as a ``$1:2$ tiling.''  This single tile is 
subdivided into five triangles that are all congruent to each
other
as shown in Figure~\ref{fig:basicPin}.

If one were to dilate the subdivision in Figure~\ref{fig:basicPin}
by a factor of $\sqrt{5}$ and then rotate and translate the resulting
figure so that the dilated copy of $C$ ended up coincident with the
original tile $P$, then a larger subset of ${\cal E}^2$ would now be tiled.
The above
subdivision scheme is then applied to each of the five
copies of $P$, and then another dilation followed by rotation and 
translation is carried out.
Continuing this process infinitely would lead to a tiling
of the plane. Thus, in the case of the standard pinwheel tilings, $P$
and $P_R$ (where $P_R$ denotes the reflection of $P$ about the x-axis)
form the set of fixed prototiles and the tiling uses
translations and rotations of this set. 

For our purposes however,
we will concentrate just on the subdivision step and omit the dilation,
translation and rotation steps, leading to the
``subdivision'' pinwheel tiling in which the cell diameter tends to zero
and the area of the plane covered by the mesh does not expand from 
step to step.  This is because we are interested in
generating a mesh with varying amounts of
refinement for a fixed region rather than a mesh that ultimately
covers ${\cal E}^2$.  In the subdivision pinwheel tiling, one starts with a fixed
$1:2$ triangle and then repeatedly subdivides first the initial
triangle and then each subtriangle into five
congruent subtriangles using the above rule.  

One can enumerate the rotation angles of the child triangles with
respect to $P$ and $P_R$ as $R_{\theta}P_R$, $R_{\pi
+\theta}P$, $R_{\theta}P$, $R_{\theta}P_R$, $R_{\frac{\pi}{2} +
\theta}P$ where $R_{\theta}$ is rotation by $\theta$ in the
counterclockwise direction. For the standard $1:2$ right triangle,
$\theta = \arctan(1/2)$ and $\theta/\pi$ in this
case is irrational. The significance of this is as follows. As the number of
subdivisions goes to infinity, so do the distinct orientations of the
triangles. For example, suppose we keep track of the orientation of
all triangles of type $C$ with respect to the parent triangle in the
subdivisions. As can be seen in Figure ~\ref{fig:basicPin}, the angle
made by a triangle of type $C$ with respect to the parent in the
$n$th subdivision is $n\theta$. Since $\theta/\pi$
is irrational, $n\theta$ will represent a different angle for each
$n$.

This presence of an infinite number of orientations
leads to a special property
known as the isoperimetric property 
\cite{Radin-isoperimetry}.
For a tiling of ${\mathcal E}^2$,
\emph{isoperimetry} means that given an
$\epsilon > 0$, there exists an $R$ such that for any two points $P$
and $Q$ on the boundaries of the triangles with
$||P-Q||>R$, the shortest path from $P$ to $Q$ 
that uses only tile edges has length at most
$(1+\epsilon)\Vert P-Q\Vert$.   Here $\Vert P-Q\Vert$
denotes the Euclidean distance from $P$ to $Q$, which will
also be denoted as $|PQ|$.

There is an analogous property for the subdivision pinwheel tiling.
In this case, let $P$, $Q$ be two points on the boundary of the
initial triangle.  Then for every $\epsilon>0$, there exists an 
$n$ such that after $n$ recursive subdivisions of the initial
triangle, the shortest path from $P$ to $Q$ using only triangle
edges is at most
$||P-Q||(1+\epsilon)$.
This theorem can be generalized so that $P$ and $Q$ do not have
to be on the boundary of the initial triangle but may be any
two distinct points.

The isoperimetric property is the reason that pinwheel tilings are
attractive for cohesive interface modeling.  
Consider a finite region $\Omega\subset {\cal E}^2$ 
tiled with an infinite sequence
of pinwheel tilings $\M_1,\M_2,\ldots$ in which
the triangles in $\M_i$ all have side lengths $h_i$, $2h_i$, $\sqrt{5}h_i$,
and $h_i\rightarrow 0$ as $i\rightarrow \infty$.
Then for an arbitrary straight segment of length $l$ connecting
$p\in\Omega$ to $q\in\Omega$,
and for an arbitrary $\epsilon>0$, there exists an $I$ such that
in each of the tilings $\M_I, \M_{I+1},\ldots$, there is a path from
$p$ to $q$ using only mesh edges (except for initial and
ending segments to connect
$p$ and $q$ to the boundaries of the triangles that contain them) such
that the length of the path is $l(1+\epsilon)$.  
\papselect{
We will give a proof of
this result in a more general setting in \sref{proof}.
}
{
We will state this result 
in a more general setting in \sref{proof}.
}

Since the above result holds for an arbitrary line segment, it also
holds for any piecewise smooth curve or network of such curves.  The
reason is that a  network of piecewise smooth curves can be approximated
arbitrarily accurately with a network of line segments.  Then each of the line
segments can be approximated arbitrarily accurately with paths
of the pinwheel tiling.

Thus, when used for cohesive fracture,
the pinwheel tiling has the property that
all possible crack paths are approximated as accurately as desired
(in terms of their length)
by paths that use only mesh edges,
as the mesh diameter tends to zero.  As we
shall see in the next section, more common mesh generation techniques
do not have this
property.

\section{A computational experiment}
\label{experiment}

In this section we carry out some simple experiments to quantify
the isoperimetric property of the 1:2 tiling.
Since our interest here is in meshes, we first explain how
to convert the 1:2 pinwheel tiling to a mesh.  It is apparent
from Figure~\ref{fig:basicPin} that the pinwheel tiling is almost
a triangulation except for the hanging node bisecting the medium
side of triangle $E$.  We define a {\em hanging node} of a 
planar subdivision into triangles to be a point that is a vertex of one
triangle but lies on the strict relative interior of an edge of
another triangle.  

It is fairly simple to make the pinwheel tiling a mesh
\cite{Radin94}: we divide every triangle into two by joining its
medium vertex to the midpoint of its medium edge.  In fact,
it is not 
necessary to split all the triangles, and in our example we have
obtained a mesh by splitting a certain subset of the tiles.
This splitting is done only on the finest level
of the pinwheel subdivision.

Our computational experiment is as follows.  Starting from a $1:2$
rectangle, we divide it into two $1:2$ triangles and then apply the
pinwheel subdivision $n$ times to each of the $1:2$ triangles.  
Thus, the final tiling has $2\cdot 5^n$ triangles.
The
resulting tiling of the original rectangle is then converted to a mesh
using the technique in the last paragraph.  

Given a tiling $\mathcal T$ of a domain $\Omega$, 
let $\Sk(\mathcal T)$ be
the 1-skeleton of $\mathcal T$, 
that is, the union of all edges of all triangles,
and let $V({\mathcal T})$ be the set of all vertices of ${\mathcal T}$.
Let $l$ be a positive parameter chosen small enough
so that $\Omega$ contains a disk of diameter $l$.
We propose to evaluate
isoperimetric quality of the triangulation with the following quantity,
which we refer to as the {\em $l$-path deviation ratio}:
\begin{eqnarray*}
\papselect{
\devi_l({\mathcal T})&=&\max\left\{\frac{\dist_{\Sk({\mathcal T})}(p,q)}
{\Vert p-q\Vert_\Omega}:
\mbox{$p,q \in V({\mathcal T})$ and $\Vert p-q\Vert\ge l$}
\right\}.
}
{
\devi_l({\mathcal T})&=&\max\left\{\frac{\dist_{\Sk({\mathcal T})}(p,q)}
{\Vert p-q\Vert}:\right. \\
& &
\mbox{$p,q \in V({\mathcal T})$ and $\Vert p-q\Vert\ge l$}
\bigg\}
}
\end{eqnarray*}
Here, 
$\dist_{\Sk({\mathcal T})}(\cdot,\cdot)$ means shortest distance among paths
restricted to $\Sk({\mathcal T})$.
The notation $\Vert p-q\Vert_\Omega$ means the geodesic distance from
$p$ to $q$, i.e., the shortest path among paths lying in $\Omega$.
Thus, this quantity measures the maximum ratio between the paths
in the mesh versus geodesic paths.  
Clearly for any mesh ${\mathcal T}$ of any polygon, 
$\devi_l({\mathcal T}) > 1$.
The pinwheel mesh of the
1:2 rectangle has the
property that for any $l\in(0,1)$, $\devi_l({\mathcal{P}\mathcal{T}}_m)\rightarrow 1$ 
as $m\rightarrow \infty$ where 
${\mathcal P}{\mathcal T}_m$ is the pinwheel tiling  of the $1:2$
rectangle after $m$ levels of refinement.

Our experiment is to evaluate $\devi_1({\mathcal P}{\mathcal T}_m)$ for 
${\mathcal P}{\mathcal T}_1,\ldots,{\mathcal P}{\mathcal T}_5$.
The results are depicted in Table~\ref{devtable}.
\papselect{
The worst-case shortest path is shown in Figure~\ref{longpath}.
}
{
}

\begin{table}
\caption{Direct computation of deviation ratios for the first five
levels of pinwheel subdivision.}
\label{devtable}
\begin{center}
\begin{tabular}{cc}
\hline 
$n$ & $\devi_1({\mathcal P}{\mathcal T}_n)$ \\
\hline
  1  &                 1.3416 \\
   2 &                 1.1948 \\
   3  &                1.1843 \\
   4  &               1.1264 \\
   5 &                 1.0831\\
\hline
\end{tabular}
\end{center}
\end{table}

\papselect{
\begin{figure}
\begin{center}
\epsfig{file=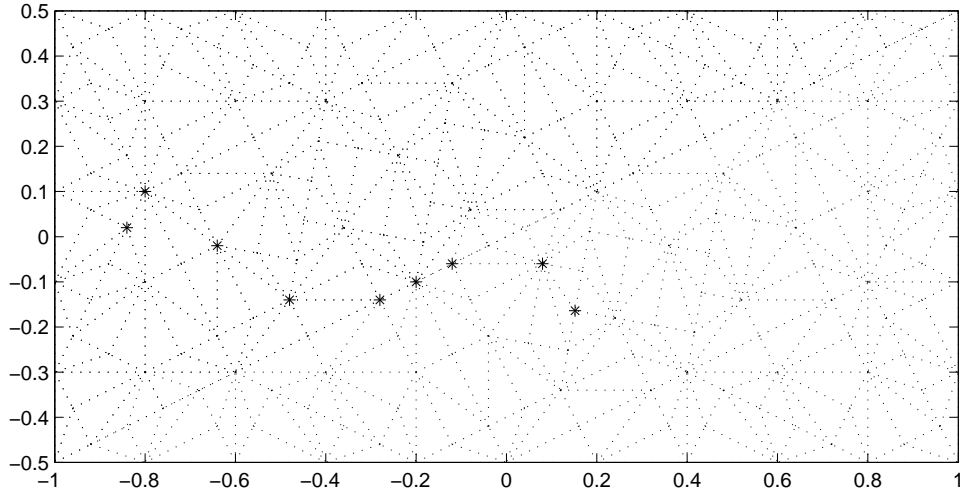,width=0.8\columnwidth}
\end{center}
\caption{The path with the largest deviation ratio
in ${\mathcal P}{\mathcal T}_4$ is marked with asterisks.}
\label{longpath}
\end{figure}
}
{
}

\begin{figure}
\begin{center}
$$
\begin{array}{cc}
\epsfig{file=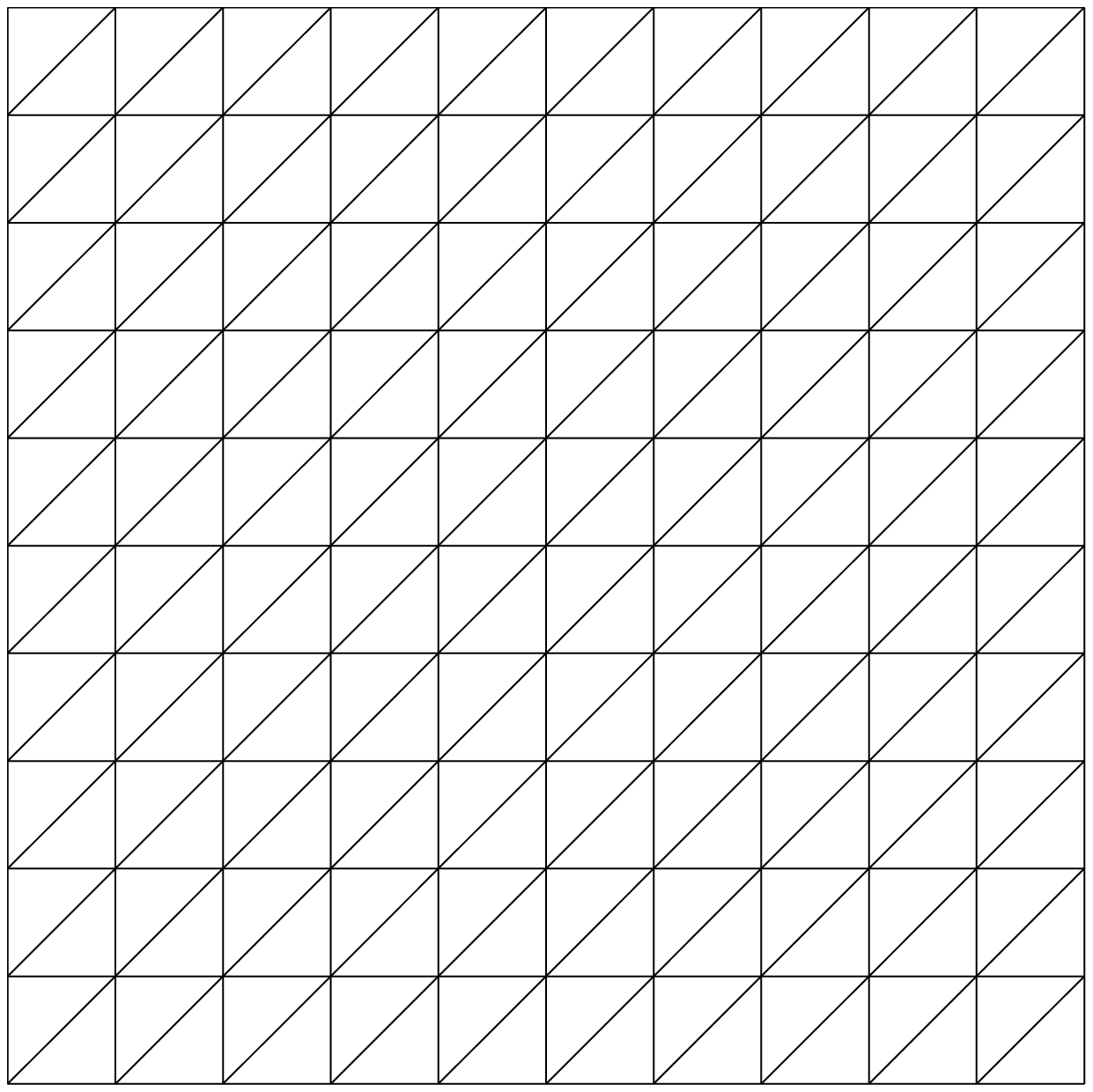,width=0.4\columnwidth} &
\epsfig{file=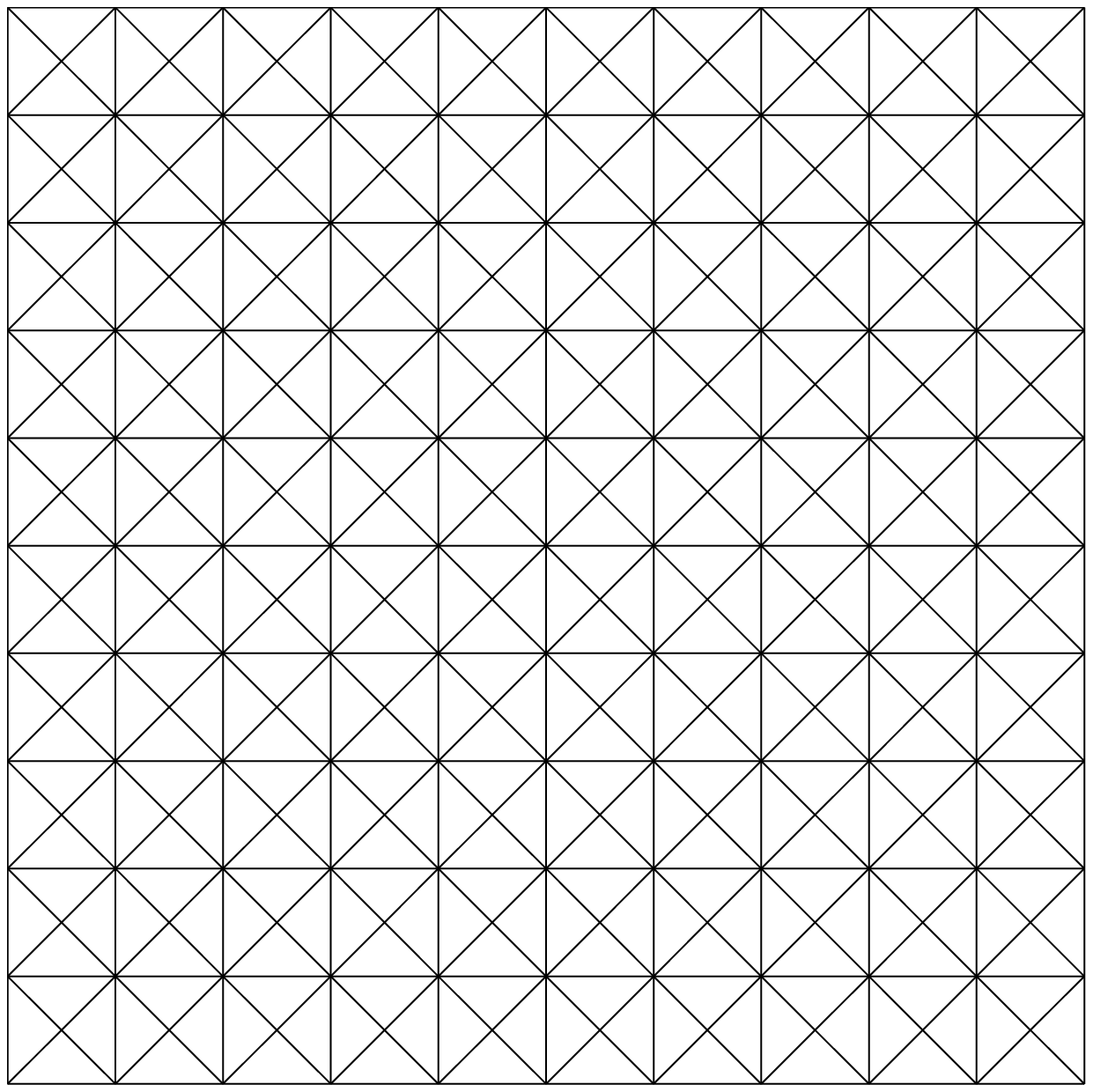,width =0.4\columnwidth} \\
(a) & (b)
\end{array}
$$
\end{center}
\caption{Two regular meshes of a square}
\label{fig:regularMesh}
\end{figure}

In contrast, consider the 
meshes in Figure~\ref{fig:regularMesh}. The
deviation ratios of these
meshes have lower bounds greater than 1
irrespective of the number of
subdivisions.  In particular, the lower bound is 
$\sqrt{2}\approx 1.414$ for
the mesh in Figure~\ref{fig:regularMesh}$(a)$.
For the mesh that was used by Xu and Needleman
\cite{Xu-Needleman} (one of the first papers on cohesive
finite element modeling), which is shown in
Figure~\ref{fig:regularMesh}$(b)$
and is sometimes called a ``cross-triangle
quadrilateral'' mesh,
the worst case deviation ratio can be
shown to be approximately equal to 
$1.082$ in the limit as
the mesh cell size tends to 0.

\section{Generalization of Pinwheel Tilings}
\label{sec:generalization}

The $1:2$ pinwheel tiling discussed up to
now was extended to a tiling with an
arbitrary right triangle and its reflection
as a prototiles by Sadun \cite{Sadun}. The
small angle of the prototile determines the finiteness of the
orientations and sizes of the tiles in the tilings that are discussed
in \cite{Sadun}. We now describe our approach to extend the pinwheel
subdivision to arbitrary (non-right) triangles.

\begin{figure}
\begin{center}
\epsfig{file=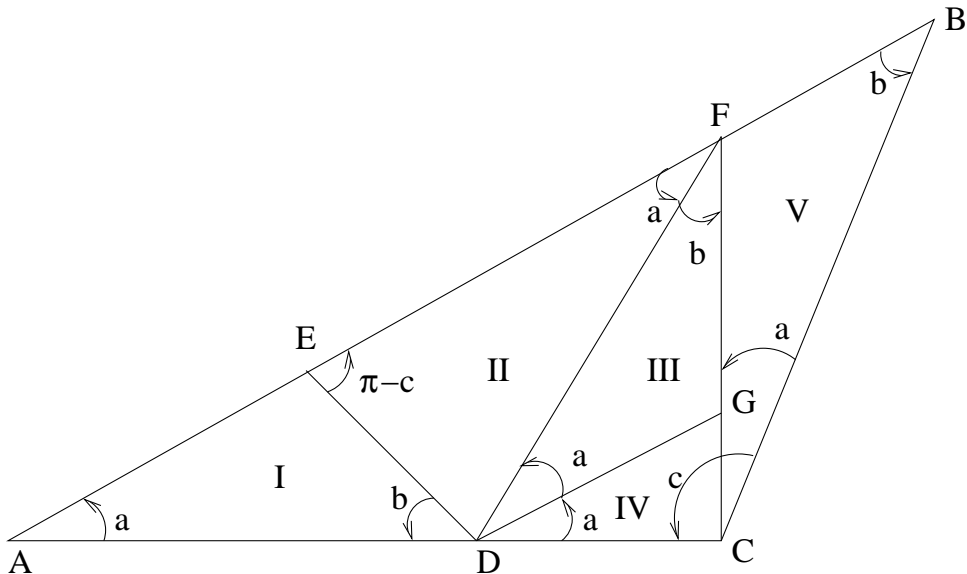, width = 0.8\columnwidth}
\end{center}
\caption{Our generalized pinwheel subdivision of an arbitary
triangle into five subtriangles.}
\label{fig:genPin}
\end{figure}

First we propose a way of subdividing a general triangle and show
that any number of subdivisions would produce triangles similar to a
finite set of prototiles. Consider the triangle shown in 
Fig.~\ref{fig:genPin}. We denote the vertices by $A$, $B$ and $C$ in
clockwise order and the
included angles at these vertices by $a$, $b$ and $c$ respectively.
Assume also $a<c$.
First, draw the segment $CF$ such that $F$
is a point on $AB$ and $\angle{FCB} = a$ measured counterclockwise
from $CB$. From $F$ draw $FD$ such that $D$ is on $AC$ and
$\angle{DFC} = b$ measured clockwise from $FC$. From $D$ draw $E$ and
$G$ such that $E$ is on $AB$ and $G$ is on $CF$ and $\angle{ADE} = b$
clockwise from $DA$ and $\angle{GDC} = a$ counterclockwise from
$DC$. Thus, we
have a subdivision of a general $\Delta{ABC}$ into five triangles of
which $I$, $III$ and $V$ are similar to the parent and the remaining
two $II$ and $IV$ are similar to each other but not to the parent.
Note that we required $a< c$ 
to make this construction but we did not require any ordering
on $b$.

\begin{theorem}
The above procedure for subdivision produces triangles with angles
belonging to the set $A_1 = \{a,b,c\}$ or to the set
$A_2=\{a,c-a,\pi-c\}$.
\label{prop:prop1}
\end{theorem}
\begin{proof}  This is obvious by simply checking all the
angles in Figure~\ref{fig:genPin} and using the fact that angles of a triangle
sum to $\pi$.
\end{proof}

\begin{theorem}
If the above subdivision procedure is used recursively on the
subtriangles, then any triangle produced has angles either from
$A_1$ or $A_2$.
\label{conju}
\end{theorem}
\begin{proof}
One checks that if we define
$a'=a$, $b'=c-a$, $c'=\pi-c$ then $\{a',c'-a',\pi-c'\}=\{a,b,c\}$.
\end{proof}

For the rest of this paper, we say that a triangle with angles
$\{a,b,c\}$ (listed in this order) is {\em conjugate} to a triangle with angles
$\{a,c-a,\pi-c\}$.  The point of
Theorem~\ref{conju} is that conjugacy is a symmetric relationship.
We remark that if the original triangle is a right triangle, i.e.,
$c=\pi/2$, then this triangle is similar to its conjugate.  This is
the case considered by \cite{Sadun}.

These two theorems imply a procedure for subdividing any
initial triangle $T_1=\bigtriangleup ABC$ with angles $a$, $b$, $c$.  Assume
$a\le b\le c$.  Apply the first subdivision rule to get five
smaller triangles.  Then, for the three similar to $T_1$, reapply
the same rule recursively.  For the two conjugates,
apply the other rule.  For the conjugate triangles,
we do not necessarily have the order $a\le c-a\le \pi-c$,
but we do not need that order.  We need only the inequality $a<\pi-c$,
which must be true since $a+b+c=\pi$.

This procedure runs
into a difficulty when $c\approx a$ (i.e., the initial triangle is
close to equilateral) because in this case the conjugate
triangle
will have a bad aspect ratio.
We get around this problem as follows.
If $c\approx a$, then  we first subdivide the initial triangle into three about
its in-center, that is, we join the in-center to the vertices of the
original triangle and form three subtriangles. We use a cutoff in our
algorithm: if $c-a$ is less than the cutoff, then the preliminary
tripartition is carried out.
The cutoff for $c-a$ is
chosen to optimize the smallest angle. In other words, a parent is
divided about the in-center if the smallest angle prior to division is
smaller than after the division. Here smallest angle happens to be the
minimum of the angles in the two sets $A_1$ and
$A_2$ for a given set of angles $\{a,b,c\}$ and can be shown
to be $ \approx 0.4$ rad. Thus, we take the cutoff to be 0.4 rad.

\section{Isoperimetric property}
\label{proof}

This section is devoted to showing the
result 
that the generalization of the pinwheel tiling introduced
in the previous section
obeys an isoperimetric inequality.  The analysis and
proof technique in this section
closely follow the proof from \cite{Radin-isoperimetry}.
The following is the key lemma in the proof of isoperimetry.

\begin{lemma}
Let triangle $T=\bigtriangleup ABC$ be as above.  Assume $a/\pi$ is an
irrational number, where $a$ is the angle of $T$ at $A$.
Let $\theta\in[0,2\pi)$ and $\epsilon>0$ be arbitrary.
Then there is a refinement
of $T$ following the above rules that contains a triangle edge
$e$ such
that the angle between $e$ and the $x$-axis lies in the interval
$(\theta-\epsilon,\theta+\epsilon)$.

Furthermore, the length of $e$ is at least $\zeta(a,b,c,\epsilon)L$,
where $a,b,c$ are the angles of $T$, $\epsilon$ is as above, $\zeta()$ is
a 
fixed positive-valued function, and $L$ is the longest side-length of $T$.
\label{allangles}
\end{lemma}
\begin{proof}
Observe that Triangle III in the above subdivision is similar to the
initial triangle $T$
but is rotated by angle $a$. Call this triangle $T'$.
If this triangle is subdivided by the
same rule again, there will be another smaller copy of $T$, say $T''$,
rotated
by $2a$ etc.  The infinite sequence $a,2a,3a,\ldots$ taken mod $2\pi$
is dense in the interval $[0,2\pi)$ by the assumption that $a/(2\pi)$
is irrational.  Therefore, for some sufficiently fine mesh, there is
an edge $e$ of triangle $T^{(k)}$
in the interval $(\theta-\epsilon,\theta+\epsilon)$.

For the second part of the lemma, observe that
for any $\epsilon>0$ there is an $n\equiv n(\epsilon,a)$ such that
every point in $[0,2\pi]$ is distance (mod $2\pi$) at most $\epsilon$
from at least one point in the set $\{a,2a,3a,\ldots,na\}$.
Therefore, one of $T,T',\ldots,T^{(n)}$ described in the last
paragraph will have the desired edge $e$.
The longest side-length of $T$ is $L$; 
the longest side-length of $T'$ is
$q(a,b,c)L$, where $q$ is some universal
function (not depending on anything other than $a,b,c$) derived
from our construction.  By similarity, the longest edge of $T^{(2)}$
has length $q(a,b,c)^2L$.
Thus, if we define 
$\zeta(a,b,c,\epsilon)\equiv \gamma(a,b,c)q(a,b,c)^{n(\epsilon,a)}$, where
$\gamma(a,b,c)$ is the ratio of the shortest to longest side length of $T$,
then the length of $e$ is at least $\zeta(a,b,c,\epsilon)L$.
This proves the second part of the lemma is also satisfied.
\end{proof}

For the first main theorem of this section, we need one more definition.
We say that a generalized tiling $\T'$ of a triangle $T$ {\em refines}
another generalized tiling $\T$ of $T$
provided that for each tile $\tau$ of $\T$,
either $\tau$ appears in $\T'$ or a subdivision of $\tau$
appears in $\T'$.  This definition implies that
$V(\T)\subset V(\T')$ and $\Sk(\T)\subset \Sk(\T')$.
The first main theorem for this section is as follows.

\begin{theorem}
Let  $T=\bigtriangleup ABC$ be a triangle
with angles $a,b,c$ such that $a<c$ and $a/\pi$ is irrational.
Let ${\mathcal T}^0,{\mathcal T}^1,\ldots$ be an infinite
sequence of generalized tilings of $T$ generated by the rules above.
For each $i$, let $y_i$ be the maximum tile diameter in ${\mathcal T}^i$.
We assume the sequence of tilings has the following two properties:
(a) ${\mathcal T}^{i+1}$ refines ${\mathcal T}^i$, and
(b) $y_i\rightarrow 0$ as $i\rightarrow \infty$.
Let $P,Q$ be any two points on the boundary of $T$.  Then
$$\lim_{i\rightarrow\infty}
\dist_{\Sk({\mathcal T}^i)}(P,Q)= |PQ|.$$
\label{isoperim1}
\end{theorem}

In other words, every straight-line
path connecting two points $(P,Q)$
on the boundary of $T$ is approximated
with arbitrary accuracy by a path of edges of the tiling.

\papselect{
\begin{proof}
In order to prove this theorem, we require a simultaneous analysis
of tilings of a conjugate triangle.  Therefore, let us change notation
so that the original triangle is $T_1$, its conjugate is $T_2$, and there
are two sequences of tilings with the above two properties, namely,
$\T_1^0,\T_1^1,\T_1^2,\ldots$, which are tilings of $T_1$, and 
$\T_2^0,\T_2^1,\T_2^2,\ldots$, which are tilings of $T_2$. 

Without loss of generality, let us further assume that $\T^0_\nu$ 
for $\nu\in\{1,2\}$ is simply $\{T_\nu\}$, and that each subsequent
$\T^i_\nu$ is obtained from $\T^{i-1}_\nu$ by splitting exactly one
tile (so that $\T^{i}_\nu$ has exact $4i+1$ tiles).
This assumption is without loss of generality because we can take our
original given sequence $\T^1_\nu,\T^2_\nu$, etc., and insert all intermediate
tilings (i.e., if the original $\T^i_\nu$ was obtained from $T^{i-1}_\nu$
via $\theta$ subdivision operations, then we can insert $\theta-1$ intermediate
tilings in the sequence).  If we prove that the limiting property holds
for the augmented sequence, then it certainly also holds for the
original sequence.

We make the following preliminary observation about generalized
pinwheel tilings.
If $S$ is any tiling of $T_\nu$  for $\nu=1$ or $2$ obtained
from the above generalized subdivision rules,
then there exists an $i$ such that $\T^i_\nu$ refines
$S$.  

We use the following additional
definitions and notation to prove the theorem.
\begin{itemize}
\item  Let $\partial T_\nu$ denote the boundary of $T_\nu$, $\nu=1,2$.
Thus  $\partial T_1$ and $\partial T_2$ are each unions of three segments.

\item
Define $X_\nu=\partial T_\nu\times \partial T_\nu$,  for $\nu=1,2.$

\item For any $p = (P,Q)\in X_\nu$, let $f_\nu(p,n)$ denote
$\dist_{\Sk(\T_\nu^n)}(P,Q)$.

\item
For $p=(P,Q)\in X_\nu$,
let
$g_\nu(p,n)=f_\nu(p,n)/\|P-Q\|$, $\nu=1,2$.
If we took the maximum of this quantity over choices of $p$, we
would arrive at a quantity analogous to the
``deviation ratio'' introduced above in Section~\ref{experiment}.
Clearly $g_\nu(p,n)\ge 1$
for all $p,n$.  If $P=Q$, then define $g_\nu(p,n)$ to be 1.
\item
Let $F_\nu(p)=\inf_{n \geq 0}{f_\nu(p,n)}$.  Note that $f_\nu(p,n)$
is a nonincreasing function of $n$ (because every edge
of $\T_\nu^n$ is covered by edges of $\T_\nu^{n+1}$
for all $n$), so this $\inf$ is also the limit of the sequence.

\item
Let $G_\nu(p)=\inf_{n \geq 0}{g_\nu(p,n)}$.  
Clearly $G_\nu(p)\ge 1$ for all $p$.  For the same reason as
above, $G$ is also the limit of the sequence.  

The main theorem
now reduces to showing that
$G_1(p)=1$ for all $p\in X_1$.  Our proof technique requires us
to claim more strongly that 
$G_\nu(p)=1$ for all $p\in X_\nu$ and for $\nu=1$ and $2$.

\item
Let $s_\nu$ be the length of the shortest edge of $T_\nu$
for $\nu=1,2$.
Let $t_\nu$ be the length of the shortest edge among the
five triangles that result from one application of the
splitting rule to $T_\nu$, and let $\rho_\nu=s_\nu/t_\nu$.

\item
Let $X'_\nu\subset X_\nu$ denote points $p=(P,Q)$ such that
$P,Q\in\partial T_\nu$ and such that $\dist_{\partial T_\nu}(P,Q) \ge t_\nu$.
Note that $X'_\nu$ is a compact set under the norm specified above.
\end{itemize}

The reason for introducing $X'_\nu$ is that $G_\nu$ is continuous on
$X'_\nu$ as the following argument shows.  Observe that 
for $p=(P,Q)\in X'_\nu$, $G_\nu(p)=F_\nu(p)/\Vert P-Q\Vert$.
The function $F_\nu$ is continuous on all of $X_\nu$
because it is a metric.
The denominator $\Vert P-Q\Vert$ 
is also continuous and bounded away from 0 on
$X'_\nu$, hence $G_\nu$ is continuous on this set.
(Once the theorem is proved, then it is established that
$G_\nu$ is continuous on all of $X_\nu$, but this is
not so easy to prove at this stage of the argument.)

The following lemma shows 
that it suffices to analyze $X'_\nu$ rather
than all of $X_\nu$.

\begin{lemma}
For $\nu=1,2$,
$$\sup\{G_\nu(p):p\in X_\nu\} 
\le \max(\sup\{G_1(p):p\in X'_1\},
\sup\{G_2(p):p\in X'_2\}).$$
\end{lemma}
\begin{proof}
Choose an arbitrary $p=(P,Q)\in X_\nu$.  This proof will show
that 
$$G_\nu(p)
\le \max(\sup\{G_1(p):p\in X'_1\},
\sup\{G_2(p):p\in X'_2\}).$$
Taking the supremum on the left will prove the result.
If $p\in X'_\nu$
then the result is immediate since that value appears in
one of the two terms in the right-hand
side.  So assume for the rest of the proof that
$p\in X_\nu-X'_\nu$.
If $P,Q$ lie on the same side of $T_\nu$, then
the left-hand side is 1 because $f_\nu(p,n)=\|P-Q\|$ for
all $n$ in this case since the boundaries of $T_\nu$ are
covered by the edges of $\T_\nu^n$ for each $n$.  Since the right-hand
side is greater than or equal to 1, the result follows immediately.

The last case is that $P,Q$ are on distinct sides (and in particular,
are not vertices of $T_\nu$).
In this case they must be less than distance
$t_\nu$ of the same vertex by definition of $X'_\nu$.  
For the rest of the proof of this lemma, consider only the $\nu=1$ case since
the $\nu=2$ case is similar.  

First, suppose that $P,Q$ are both within
distance $t_1$ of $A$, the
vertex whose angle is $a$.  Without loss of generality, $P$ lies
on $AB$ and $Q$ lies on $AC$.
Consider the
sequence of tiles $H_0=T_1,H_1,H_2,\ldots$ such that $H_i$ is the tile
from $\T_\nu^i$ that contains vertex $A$.
Each of these tiles
is similar to $T_1$.
The diameter of the $H_i$'s tends to 0 as $i\rightarrow\infty$ by
assumption.
Thus, there is a $K$ such that
$H_K$ contains both $P$ and $Q$ but $H_{K+1}$ fails to contain one or
both of $P$ or $Q$.  Let $u$ be the length of the shortest side of $H_K$.  We
claim that either $\Vert A-P\Vert\ge u/\rho_1$ or $\Vert A-Q\Vert\ge u/\rho_1$.
The reason is that if both $\Vert A-P\Vert< u/\rho_1$ and 
$\Vert A-Q\Vert<u/\rho_1$
then $P,Q$ would both lie on the boundaries of $H_{K+1}$  since 
the side lengths of $H_{K+1}$ are all at least $u/\rho_1$ by definition
of $\rho_1$.  This would contradict the choice of $K$.

As mentioned above, $H_K$ is similar to $T_1$, and the constant
of proportionality is $u/s_1$.  Note that $H_K$ could be either 
a dilation of $T_1$ with no reflection  or a dilation
of $T_1$ with a reflection.  Assume the former
case since the latter is similar.
There exist $\bar{P}$ and $\bar{Q}$ lying on sides
$AB$, $AC$ 
of $T_1$ whose positions with respect to $AB$, $AC$ are
proportional to the positions of $P,Q$ with respect to the two
sides of $H_K$.  Since either $\Vert A-P\Vert\ge u/\rho_1$ or
$\Vert A-Q\Vert \ge u/\rho_1$ and the scaling factor between $H_K$ and $T_1$
is $u/s_1$, this means that at least one of $\bar P$, $\bar Q$ is distance
from $A$ greater than or equal to $(u/\rho_1)/(u/s_1)=t_1$.
Hence $\bar p = (\bar {P},\bar{Q})\in X_1'$.  For an arbitrary
$n>0$, consider the tiling $\T_*^n$ of $H_K$ that is obtained by shrinking
$\T_1^n$ by a factor of $u/s_1$ and translating it so that it lies
on top of $H_K$. The shortest path between $P$ and $Q$ in this tiling
is $f_1((\bar{P},\bar{Q}),n)u/s_1$ by scaling.
Also, there is an $n'$ such that the portion
of $\T_1^{n'}$ lying in $H_K$ is strictly a refinement of $\T_*^n$
by the observation made at the beginning of the proof.
The distance between $P$ and $Q$ in this tiling is $f_1((P,Q),n')$, and
since $\T_1^{n'}$ refines $\T_*^n$, 
$f_1((P,Q),n')\le f_1((\bar{P},\bar{Q}),n)u/s_1$.  Note that
$u/s_1=\Vert P-Q\Vert/\Vert\bar P -\bar Q\Vert$ by similarity,
so the previous inequality implies 
$g_1((P,Q),n')\le g_1((\bar P,\bar Q),n)$.
Take the infimum over all $n$ of both sides to conclude that
$G_1(p)\le G_1(\bar{p})$, 
thus establishing the lemma in this case.

In case that $P,Q$ are both within distance $t_1$ from vertex
$B$ whose angle is $b$, the lemma follows by the same argument since
the triangles containing $B$ in all subdivisions of $T_1$ are similar
to $T_1$.

The last case is that $P,Q$ are both within distance $t_1$
of vertex $C$ whose angle is $c$.  Say, e.g., that
$P$ lies on $AC$ and $Q$ on $BC$.
In this case, the argument is
slightly more complicated since there are two triangles containing
$C$ in the next level of subdivision.  Let $CF$ be the segment
that is the common boundary to the two triangles of the next level
of subdivision that meet vertex $C$. (Refer to
Fig.~\ref{fig:genPin}.)
Let $R$ be
the point where segment $PQ$ crosses edge $CF$.
Then the argument above shows that 
the infimum over $n$ of the
distance between $P$ and $R$ using edges from $\T_1^n$  is less than or equal
to $|PR|\cdot G_2(\bar P,\bar R)$, where $(\bar P,\bar R)\in X'_2$.
Similarly, the infimum over
$n$ of the distance between $R$ and $Q$ 
using edges from $\T_1^n$ is less than or equal
to $|RQ|\cdot G_1(\bar R,\bar Q)$, where $(\bar R,\bar Q)\in X'_1$.
Therefore, $F_1(P,Q)\le \|PQ\|\cdot \max(\sup\{G_1(p):p\in X'_1\},
\sup\{G_2(p):p\in X'_2\})$ so the result follows.
\end{proof}

Finally, we conclude the proof of the main theorem by showing
that $\sup\{G_\nu(p): p\in X_\nu'\}=1$ for $\nu=1,2$.  To this end,
choose $\nu$ (either 1 or 2)
so that $\sup\{G_\nu(p):p\in X_\nu'\}\ge
\sup\{G_{3-\nu}(p):p\in X_{3-\nu}'\}$.  Without loss of generality,
say $\nu=1$ is chosen.

Since $X'_1$ is a compact set and $G_1$ is continuous on this
set, there exists a $p=(P,Q)$ in $X_1'$ that maximizes $G_1(p)$.
If $P,Q$ lie on the same side of $T_1$, then $G_1(p)=1$ so the proof
is finished.  
Else let the corresponding pair of points where the supremum is
achieved be $p^* = (P^*,Q^*)$ and $H^*$ be the line segment joining them.
Assume (for a contradiction) that $G_1(p^*)=S>1$.

Choose $N_1$
large enough so that there exists a
tile $T'$ in $\T_1^{N_1}$ such that $T'\cap H^*$ has positive length
and is contained in the middle third $H^*$.
Let the longest edge of $T'$ be
$u_1$. For reasons to be explained below, we also choose $N_1$
large enough so that
\begin{equation}
 u_1 < 
\frac{(\sqrt{S}-1)\zeta(a,b,c,\arccos(1/\sqrt{S}))}{6S}|P^*Q^*| \label{uassump}
\end{equation}
where $\zeta()$ is the function defined by Lemma~\ref{allangles}.
Continue splitting until we reach split number
$N_2\ge N_1$ so that within
$\T_1^{N_2}$, there 
exists a tile $T''$ in $\T_1^{N_2}$ lying
inside in $T'$ that has an
edge making an angle $\theta$ with $H^*$, with $\cos{\theta} >
1/\sqrt{S}$.  
This is possible by Lemma~\ref{allangles}.
See Figure~\ref{prooffig}.
Let $L_2$ be the length of this edge.  By
the second part of the
lemma, we may assume $L_2\ge \zeta(a,b,c,\eta)u_1$, where
$\eta=\arccos(1/\sqrt{S})$.

\begin{figure}
\begin{center}
\epsfig{file=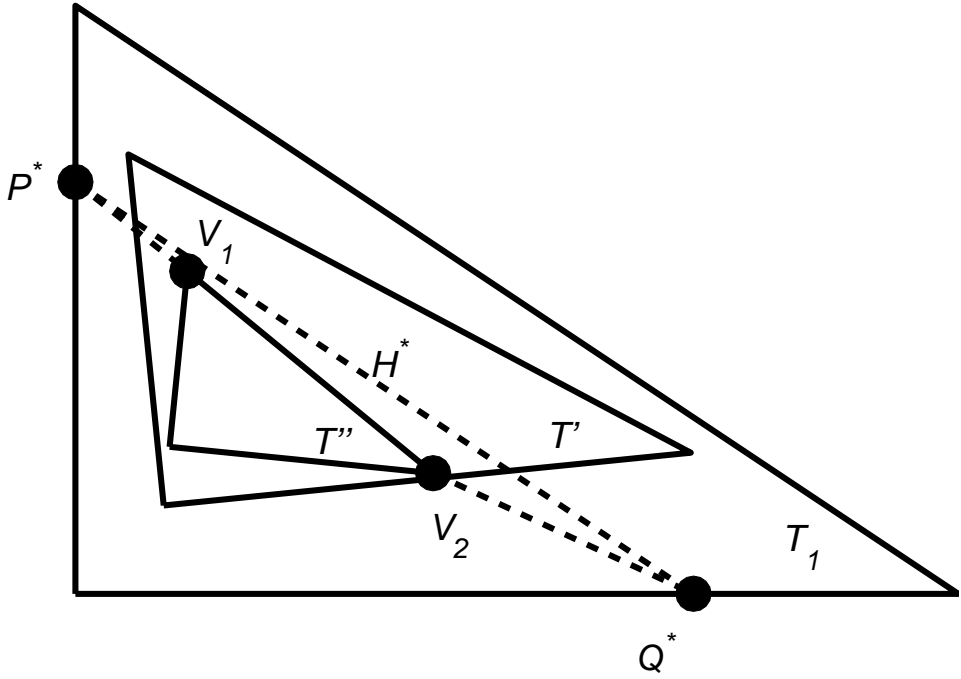,width=0.8\columnwidth}
\end{center}
\caption{Construction for the proof of Theorem~\protect\ref{isoperim1}.}
\label{prooffig}
\end{figure}

Let $V_1,V_2$ be the endpoints of this edge with 
$V_1$
being the vertex near $P^*$ and $V_2$ near $Q^*$. Observe
that  $P^*V_1$,
$V_1V_2$ and $V_2Q^*$ make up a three-segment path from $P^*$
to $Q^*$. The length of this path is $L_1+L_2+L_3$, where
$L_1=|P^*V_1|$ and
$L_3=|V_2Q^*|.$ 
(We have already defined  $L_2=|V_1V_2|$.)
Let $l_1$, $l_2$, and $l_3$ be the lengths of
the projections of $P^*V_1$, $V_1V_2$, $V_2Q^*$ respectively
onto $H^*$.  
Because $V_1$ lies within $T'$ while $H^*$ crosses through $T'$, the
distance from $V_1$ to $H^*$ is at most $u_1$, hence
\begin{eqnarray*}
L_1 &\leq& \sqrt{l_1^2+u_1^2} = l_1\sqrt{1+u_1^2/{l_1^2}} \\
&\leq & l_1\left(1+u_1^2/l_1^2\right) \\
& =& l_1 + \frac{u_1^2}{l_1} 
\end{eqnarray*}
hence
\begin{equation} 
L_1 -l_1 \leq \frac{u_1^2}{l_1} \leq \frac{3u_1^2}{|P^*Q^*|}. \label{eq:gen.eq1}
\end{equation}
(The factor of 3 arises because $\Vert P^*-V_1\Vert \ge \Vert P^*-Q^*\Vert/3$
as assumed earlier.)
Similarly,
\begin{equation}
L_3 - l_3 \leq \frac{3u_1^2}{|P^*Q^*|}. \label{eq:gen.eq2} 
\end{equation}
Next, because
$\cos{\theta} > \frac{1}{\sqrt{S}}$
where $\theta$ is the angle between $H^*$ and $V_1V_2$, 
we have
$\sqrt{S}l_2 > L_2$.
Therefore, 
\begin{equation}
Sl_2 - L_2 > (\sqrt{S}-1)L_2.  \label {eq:gen.eq3}
\end{equation}

Next, note that $F_1((P^*,Q^*))\le SL_1+L_2+SL_3$ thanks
to the existence of the three-edge path $P^*V_1V_2Q^*$. 
The reasoning is as follows.
From $P^*$ to $V_1$ there is a straight-line path of length $L_1$.
This path cuts through a finite list of triangles, say $\phi$
triangles,  within the tiling
$\T_1^{N_2}$, since $P^*$ and $V_1$ both lie on triangle
edges of this tiling.  Let the individual segments within these triangles
be of length $p_1,\ldots,p_\phi$.  By construction, these quantities sum
to $L_1$.  Then by further refinement, we can find
paths within the tiling with lengths less than or arbitrarily close to
$p_1S$, $p_2S$, etc.\ since $S$ is the
factor that is the
maximum amount longer that an edge path  in 
refinements of either ${T}_1$ or ${T}_2$ can be versus the
straight-line path.
So the infimum
of the lengths of these paths added up is at most $SL_1$.  The same
reasoning accounts for the term $SL_3$.  Finally, the edge $V_1V_2$ is
length $L_2$ and is already in the tiling.

Use (\ref{eq:gen.eq1}), (\ref{eq:gen.eq2}), (\ref{eq:gen.eq3})
and the equality $|P^*Q^*|=l_1+l_2+l_3$ to
bound $SL_1+L_2+SL_3$:
\begin{eqnarray}
F_1((P^*,Q^*)) &\le& SL_1 + SL_3 + L_2\nonumber\\
&=& S(L_1-l_1)+S(L_3-l_3)-(Sl_2-L_2)+S(l_1+l_2+l_3) \nonumber \\
  & <& 6u_1^2\frac{S}{|P^*Q^*|} - (\sqrt{S}-1)L_2 + S|P^*Q^*|.
 \label{eq:gen.eq4}
\end{eqnarray}
Multiply (\ref{uassump}) by $u_1$ on both sides and use the fact
that $L_2\ge \zeta(a,b,c,\eta)u_1$ to obtain
\begin{equation}
 u_1^2 < \frac{(\sqrt{S}-1)L_2}{6S}|P^*Q^*| \label{eq:gen.eq5}.
\end{equation}
Substituting (\ref{eq:gen.eq5}) in (\ref{eq:gen.eq4})  shows that
$F_1((P^*,Q^*)) < S|P^*Q^*|$.  But this is a contradiction, because the
hypothesis of this analysis was that $G_1((P^*,Q^*))=S$, i.e.,
$F_1((P^*,Q^*))=S|P^*Q^*|$.
\end{proof}
}
{
Due to space limitations, we defer the proof of this theorem
to the full paper.  A draft of the full paper is available from arxiv.org.
}

The preceding theorem has the drawback that it pertains only to paths
starting and ending on the boundary of the root triangle.  For
isoperimetry, we would like to generalize the result to paths with
arbitrary interior $P$ and $Q$.  Since the nodes of the
pinwheel tiling are dense in the interior (in the limit as the mesh
size is refined), the following theorem provides a suitable
generalization and will be taken as our definition of the isoperimetric
property.

\begin{theorem}
Let ${\mathcal T}^0,{\mathcal T}^1,\ldots$ be a sequence of generalized
pinwheel tilings of $T$ (satisfying $a<c$ and $a/\pi$ is irrational
as in the previous theorem) such that the maximum cell diameter tends to zero
and such that $\T^{i+1}$ refines $\T^i$
for all $i=0,1,2,\ldots$.
Let $P,Q$ be any pair of distinct points lying on $\Sk({\mathcal T}^n)$
for some $n$.  Then
$$\lim_{
\renewcommand\arraystretch{0.5}
\begin{array}{c}
\scriptstyle m\rightarrow\infty \\
\scriptstyle m\ge n
\end{array}
}
\dist_{\Sk({\mathcal T}^m)}(P,Q)=\Vert P-Q\Vert.$$
\label{isoperim2}
\end{theorem}
\begin{proof}
Consider the segment $PQ$ lying in $T$.  
Let $\epsilon>0$ be given.
Make a list $U_1,\ldots, U_r$ of
tiles in ${\mathcal T}_n$ traversed by this segment.
Since $PQ$ crosses $U_i$, define $P_iQ_i$ to be $U_i\cap PQ$.
Observe that $P_i,Q_i$ both lie on the boundary of $U_i$.
By the preceding theorem, after a sufficient number 
of further subdivisions (say $s$),
there exists a path in $\Sk({\mathcal T}^{n+s})$ between $P_i$ and $Q_i$
of length $|P_iQ_i|(1+\epsilon)$.  This choice of $s$ depends on
$i$, so take the maximum such value of $s$ (maximum over all $i=1,\ldots,r$).  
Then there is a path in $\Sk({\mathcal T}^{n+s})$ from $P$ to $Q$ of length
at most 
$$|P_1Q_1|(1+\epsilon) + |P_2Q_2|(1+\epsilon) +\cdots+|P_rQ_r|(1+\epsilon),$$
i.e., at most
$|PQ|(1+\epsilon)$.
\end{proof}

\section{Meshing an arbitrary region}
\label{algo}

\begin{figure}
\begin{center}
\epsfig{file=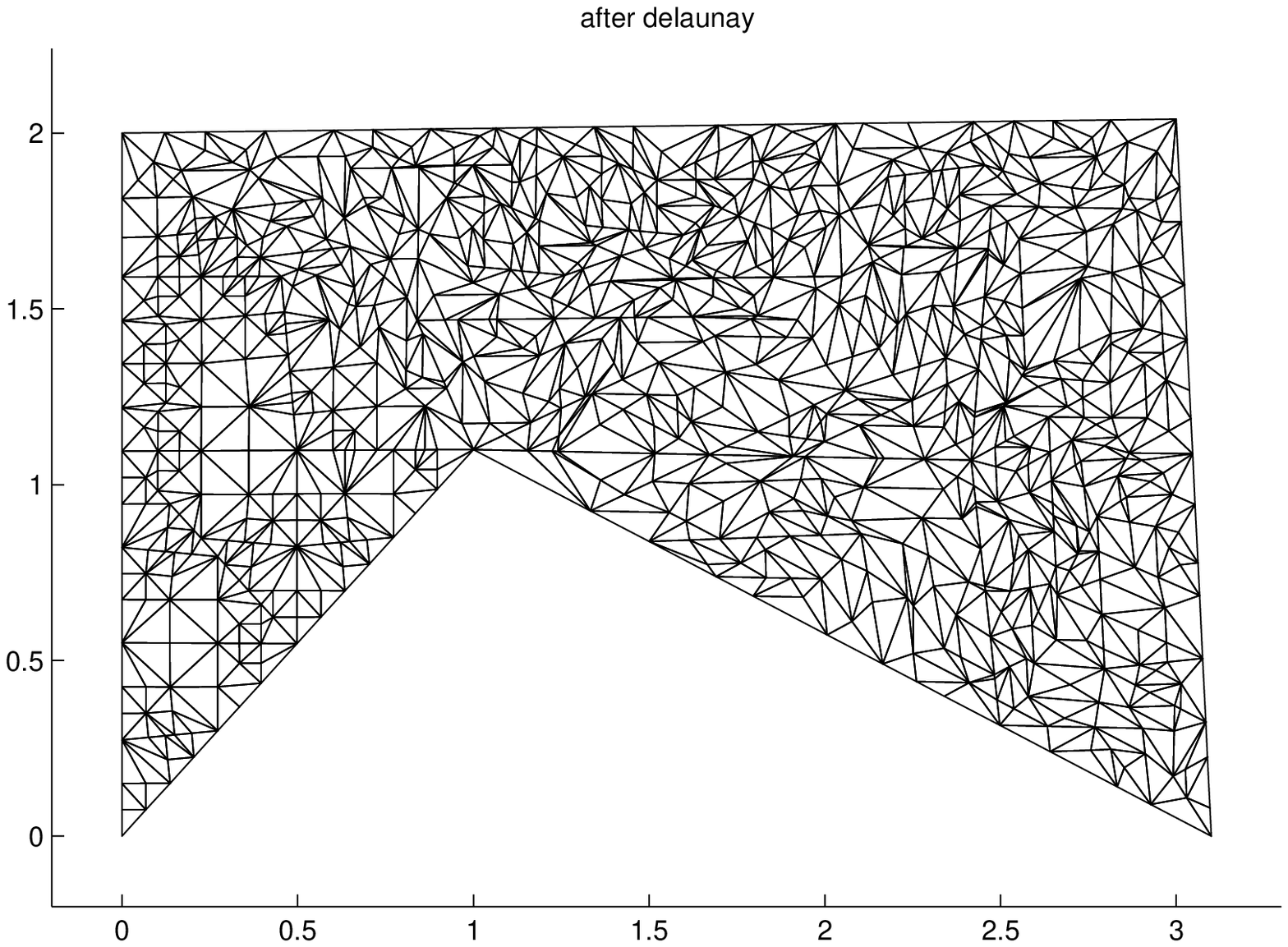, width = 0.8\columnwidth}
\end{center}
\caption{
The coarse mesh for this figure had three triangles. 
The final mesh after pinwheel tiling, collapse-node operations
and Delaunay triangulation is shown.}
\label{delaunay}
\end{figure}

In this section we present our algorithm PINW 
to mesh a region $\Omega$ with arbitrary
polygonal boundary.  A summary of PINW appears in
Figure~\ref{algpinw}.  The steps in 
this summary are described in more detail in the
remainder of this section. The current version of PINW is 1.0
and has been coded in Matlab.
An example output from this algorithm is shown in Fig.~\ref{delaunay}.

\begin{figure}
\begin{center}
\framebox{
\begin{minipage}{0.9\columnwidth}
\begin{center}
{\bf Algorithm PINW 1.0}
\end{center}

\papselect{
\begin{enumerate}
}
{
\begin{list}{\arabic{enumi}.}{\usecounter{enumi}
  \setlength{\labelwidth}{4em}\setlength{\parsep}{0pt}
  \setlength{\itemsep}{0.3ex}}
}
\item
Generate a mesh for $\Omega$ with bounded aspect ratio using
Triangle.
\item
Split triangles too close to equilateral at their in-centers.
\item
Split triangles whose smallest angle is a rational
multiple of $\pi$ at a point near the in-center.
\item
Let the set of triangles
obtained after steps 1--3 be called ${\mathcal T}_0$.
\item
Initialize a heap containing triangles that need splitting.  The
triangles are ordered so that the one whose minimum altitude is
maximum is at the top of the heap.  Initially the heap contains
all triangles from ${\mathcal T}_0$.
\item
Repeatedly remove a triangle from the heap and split it into five,
until the size of the top element of the heap is sufficiently small
according to the user's specification.  
\item
Let ${\mathcal T}_*$ be the set of tiles including those in
${\mathcal T}_0$
and all their descendants obtained by subdivision.
Let ${\mathcal T}_f\subset {\mathcal T}_*$ be the set of leaf 
tiles.
\item
Loop over all tiles in ${\mathcal T}_*$
starting from the coarsest
to determine the value
of $\big(e)$ for each edge $e$ of any tile.
\item
For each big edge (i.e., each edge in the image of the ``big'' operator), 
select one side as {\em moving} and the other as
{\em staying}. Sort the list of nodes lying on the staying
side of each such edge.
\item
Loop over tiles in ${\mathcal T}_*$ starting from the coarsest
excluding ${\mathcal T}_f$.
For each such tile $T$ and for each of its vertices $D,E,F$
as labeled in Figure~\ref{fig:genPin},
let $e$ be the maximal big edge containing the particular vertex.
If this vertex $D$, $E$ or $F$ is on the moving side of $e$ and is very
close to a vertex $v'$ on the staying side, then displace it
to coincide with $v'$
and apply the induced affine transformation to subtriangles of $T$.
\item
Apply Delaunay triangulation to each distorted, subdivided leaf tile.
(The distortion of the leaf tiles is due to
the affine transformations in the previous
step.  The subdivision of the edges is
due to the presence of hanging nodes.)  
The collection of triangles
output from this step is a simplicial mesh of $\Omega$.
\papselect{
\end{enumerate}
}
{
\end{list}
}
\end{minipage}
}
\end{center}
\caption{Overview of the steps of the PINW algorithm.}
\label{algpinw}
\end{figure}

We first start with a coarse triangulation of the
domain.  We use the \emph{Triangle} package \cite{Shewchuk} developed by 
J.~Shewchuk, which uses Delaunay triangulation.
The triangles produced have bounded aspect
ratio.  The second preliminary step, as mentioned in \sref{sec:generalization},
locates triangles
too close to equilateral and splits them at their in-center.

\papselect{
A third preliminary step is to identify
and split triangles whose smallest angle $a$ is a rational
multiple of $\pi$.  
(As noted above, the proof of isoperimetry
requires that $a/\pi$ be irrational.)  In principle, this test could be
conducted exactly using number-theoretic methods since the coordinates
of the vertices of
each triangle, being floating-points numbers, are rational
numbers and can be treated with integer algorithms by
clearing common denominators.  Modifying a triangle in which
$a$ is a rational multiple of $\pi$ is trivial in principle
because any small random perturbation of a node of such a triangle will
lead to an angle that is not a rational multiple of $\pi$ with probability
1.

In practice, this exact test and solution are both undesirable.  
For practical use of the algorithm,
we would like to avoid the case when $a$ is close to 
a rational multiple of $\pi$ of the form $m\pi/n$ where $n$ is a small
integer.  The reason is that the presence of such a triangle in
which $a/\pi$ is irrational but is close to $m/n$ implies that,
although the isoperimetric property is asymptotically valid, the
available angles will be badly distributed (clustered around multiples of
$\pi/n$)
for modest levels of refinement.

Therefore, a more practical
heuristic is to check each smallest angle against a finite list of the form
$m\pi/n$, where $m,n$ range over a pre-selected set of small integers.
If a triangle's smallest angle comes too close to a member of this list,
then the triangle is either split into three
using a point near its in-center or is perturbed.  (The
exact in-center obviously should not be used since this would replace each
angle by half its previous value, and hence still close to a rational
multiple of $\pi$.)   This step has
not been implemented in the current version of our
code PINW 1.0 because we are still seeking the best
practical heuristic.
(Indeed, in Figure~\ref{delaunay},
one coarse triangle is close to a 45-degree right triangle, and hence one
part of the subdivision exhibits a shortage of possible directions.)
}
{
A third preliminary step is to identify
and split triangles whose smallest angle $a$ is a rational
multiple of $\pi$.  
In principle, this test could be
conducted exactly using number-theoretic methods since the coordinates
of the vertices of
each triangle, being floating-points numbers, are rational
numbers and can be treated with integer algorithms by
clearing common denominators.  Modifying a triangle in which
$a$ is a rational multiple of $\pi$ is trivial in principle
because any small random perturbation of a node of such a triangle will
lead to an angle that is not a rational multiple of $\pi$ with probability
1.

In practice, this exact test and solution are both undesirable.  
For practical use of the algorithm,
we would like to avoid the case when $a$ is close to 
a rational multiple of $\pi$ of the form $m\pi/n$ where $n$ is a small
integer. 
Therefore, a more practical
heuristic is to check each smallest angle against a finite list of the form
$m\pi/n$, where $m,n$ range over a pre-selected set of small integers.
This step has
not been implemented in the current version of our
code PINW 1.0 because we are still seeking the best
practical heuristic.
(Indeed, in Figure~\ref{delaunay},
one coarse triangle is close to a 45-degree right triangle, and hence one
part of the subdivision exhibits a shortage of possible directions.)
}

Let ${\mathcal T}_0$ be the list of triangles that are produced by these
preliminary steps.   Thus, the triangles in ${\mathcal T}_0$ form
a simplicial triangulation of the input set $\Omega$.
We call these triangles the {\em root tiles}.
The generalized pinwheel subdivision is then performed on the
the root tiles to obtain a refined tiling.
The procedure to refine the mesh used in PINW 1.0 is based on a 
simple heap \cite{AHU}. The heap is initialized with all triangles
in ${\mathcal T}_0$, which are
ordered in the heap according
to length of the minimum altitude.  The main loop for the subdivision is to
remove the top member of the heap (i.e., the unsubdivided tile with
the largest value of minimum altitude) and replace it with its five children.
The procedure terminates when the top triangle in the heap is
smaller than the user-specified mesh size requirement.

Note that during the subdivision procedure, the angles $a,b,c$ in
Figure~\ref{fig:genPin} are assigned to smallest, middle and largest
angles respectively for tiles similar to
root tiles.  For the conjugate tiles, angles $a,b,c$
are assigned according to the conjugacy relationship.  In other words,
if the angles of the root tile are $a',b',c'$ in that order, then
the angles $a,b,c$ in the conjugate tile are assigned in the order
$a=a'$, $b=c'-a'$, and $c=\pi-c'$.  This ensures that the conjugate
of the conjugate is again similar to the root tile.  

From this description, it is apparent that
PINW 1.0 supports a single global user-specified mesh size requirement.
For many applications of mesh generation, it is useful to have a finer mesh
in one part of the domain versus another.  This can also be implemented
in the framework of generalized pinwheel subdivision but is not available
in PINW 1.0.  In addition, several aspects of our analysis
that follows below would
have to be generalized to cover graded meshes.

Once the subdivision procedure is complete, the resulting tiling must
be converted to a simplicial
mesh.  For the 1:2 pinwheel triangulation, this step
is quite straightforward as mentioned in \sref{experiment}.  In the
generalized case, however, it is much more complicated and involves
several steps that we shall now describe. 

Let ${\mathcal T}_*$ be the list of all tiles in the hierarchy:
it includes the tiles in 
${\mathcal T}_0$ and all their descendants from
the subdivision procedure.  The tiles in ${\mathcal T}_*$ naturally
have a forest structure associated with them in which the forest
roots are root tiles.
Let {\em leaf tile} denote a triangle in ${\mathcal T}_*$
that is not
further subdivided during the generalized pinwheel subdivision phase. Let
${\mathcal T}_f$ be the set of leaf tiles.

The first step in converting the tiling to a mesh is to identify
for each edge $e$ of each tile $T\in{\mathcal T}_*$ the edge that we denote
$\big(e)$.  This is defined to be the edge $e'$ of a triangle $T'$
higher up in the subdivision hierarchy 
(i.e., $T$ is derived from $T'$ via a sequence
of zero or more subdivision operations)
such that $e\subset e'$, and such that $e'$ is maximal with this
property (i.e., there is no other ancestor of $T$
with an edge $e''$ that strictly contains $e'$).

For each triangle in ${\mathcal T}_0$,
$\big(e)=e$.   For some other tile $T$ with an
edge $e$, it is a straightforward
matter based on a checking a finite number of cases whether
$\big(e)=e$ or $\big(e)\ne e$.  In the latter case, $\big(e)$ can
be determined from the immediate parent of $T$ (assuming $\big(e)$
is already tabulated for the the parent's edges).  Thus, it
is possible to determine $\big(e)$ for each edge of
each tile in ${\mathcal T}_*$ with
a constant number of operations per tile.

Next, for each ``big'' edge $e$ (that is, an edge such that
 $\big(e)=e$), identify
a {\em moving} and {\em staying} side.  This choice can be quite
arbitrary, except for two stipulations. An edge $e$ adjacent on
the exterior boundary of $\Omega$ should have its inside labeled
staying (i.e., no tiles lie on its moving side).  
An edge in correspondence with $CF$ in Figure~\ref{fig:genPin}
(every big edge generated during the subdivision procedure is in
correspondence with either $DE$, $DF$, $CF$ or $DG$) should
have the side facing vertex $B$ labeled as moving.
We now identify all the nodes on the
staying side  of $e$
and sort them in order of occurrence on the
edge.   This sorted list is saved for the next phase of the
algorithm.

In the next phase, 
we loop over triangles in ${\mathcal T}_*-{\mathcal T}_f$ 
starting from the coarsest and perform
collapse-node operations on each.  Let $T$ be a tile in ${\mathcal T}_*-{\mathcal T}_f$.
Let the four vertices of $T$ introduced when it is subdivided be labeled
$D,E,F,G$ as in Figure~\ref{fig:genPin}.  We perform no operation
for $G$ since it is on the staying side of edge $CF$.
The maximal big edge containing $D$ is $\big(AC)$; call this $b(D)$.
The maximal big edge containing $E$ and $F$ is $\big(AB)$; call this
$b(E)$ and also call it $b(F)$.
Let $v$ be one of $D,E,F$. We check whether $v$ is on the moving side
of $b(v)$.  If it is on the staying side, then no
further operation is performed.  If it is on the moving side,
then we find the vertex $v'$ taken from the
staying side of $b(v)$ that is closest to $v$.
This $v'$ can be found efficiently using binary search on the precomputed
sorted lists.
If $\Vert v-v'\Vert\le \delta$,
we collapse nodes $v$ and $v'$  by displacing $v$ to $v'$.  
Here $\delta$ is a tolerance discussed more below.

\begin{figure}
\begin{center}
$$
\begin{array}{cc}
\epsfig{file=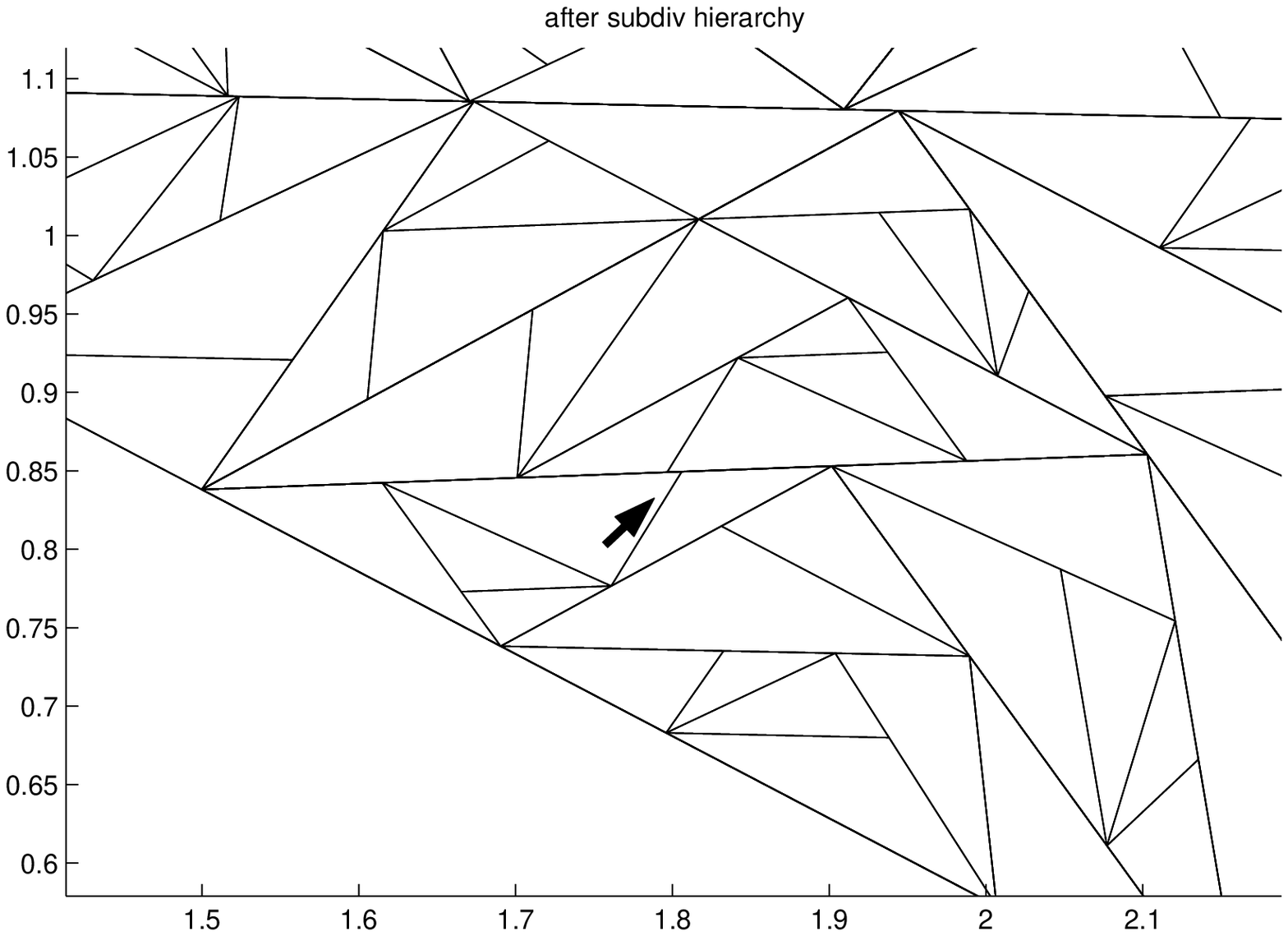,width=0.4\columnwidth} &
\epsfig{file=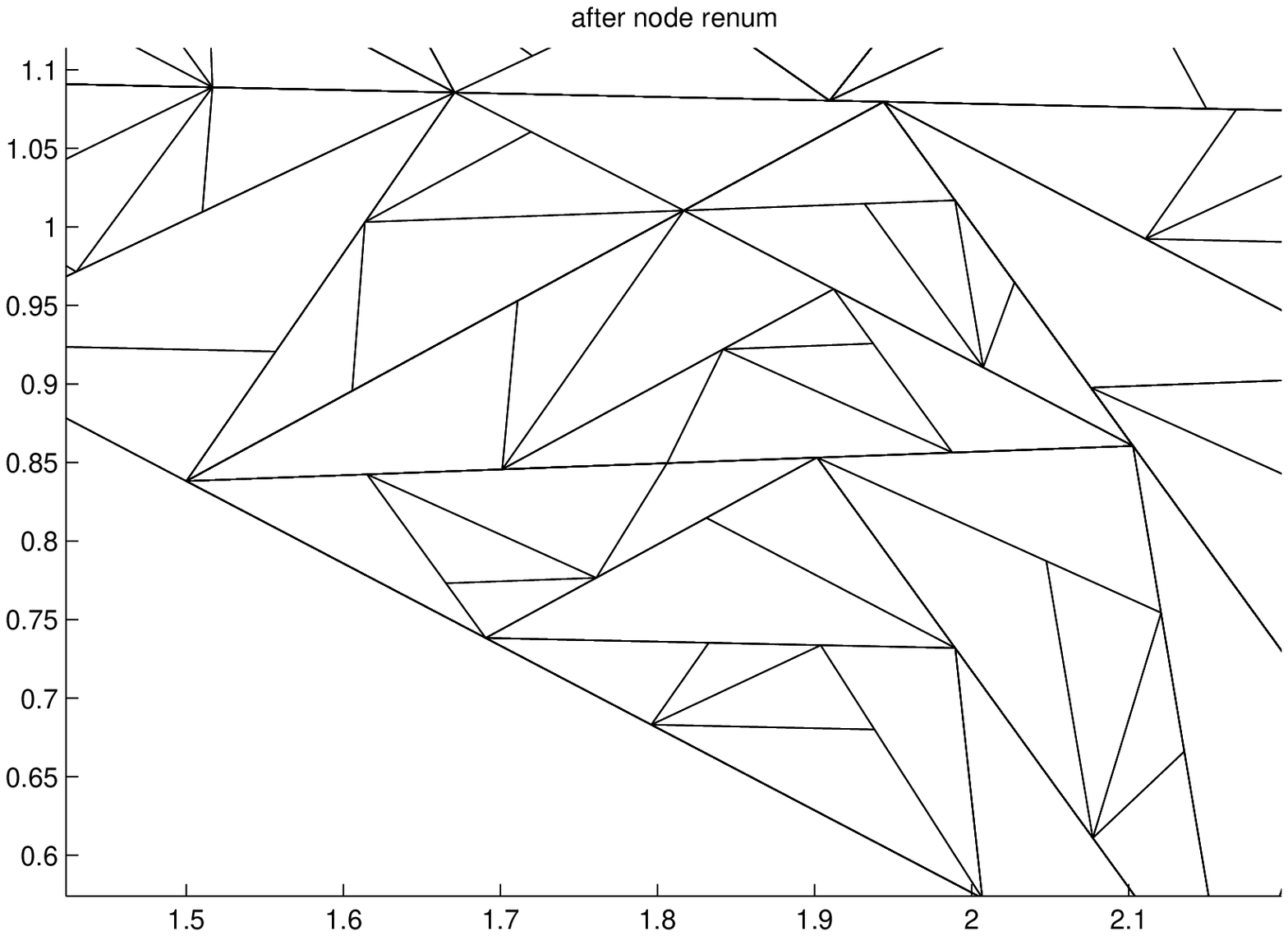,width=0.4\columnwidth} \\
(a) &
(b)
\end{array}
$$
\end{center}
\caption{Example of a collapse-node operation is shown.
A node on one side of a ``big'' edge that lies within the tolerance
of a node on the other side
is moved and merged with the nearby node on the other side.}
\label{collapse}
\end{figure}

This
displacement induces uniquely determined affine transformations on 
triangles contained in $T$ as follows.
If $v$ is
the vertex labeled $D$ in Figure~\ref{fig:genPin}, then there is
a unique affine transformation on $\bigtriangleup ADF$ that leaves
$A$ and $F$ fixed and moves $D$ to $v'$.  A second affine transformation
of $\bigtriangleup CDF$ leaves $F$ and $C$ fixed and moves $D$ to $v'$.
If $v$ is the vertex labeled $E$, then there are unique affine transformations
determined for $\bigtriangleup ADE$ and $\bigtriangleup DEF$.
Finally, if $v$ is the vertex labeled $F$, then there are transformations
for each of 
$\bigtriangleup DEF$, $\bigtriangleup CDF$ and $\bigtriangleup BCF$.
The algorithm applies all the relevant affine transformations caused
by motion of the node.  
Note that the affine
transformations agree on the boundaries between these triangles, so there
is no consistency issue regarding which transformation to apply.
These transformations move the triangle, including every node
at deeper levels of the hierarchy contained in it.  This concludes the
description of the {\em collapse-node} operation.
See Figure~\ref{collapse}
for an illustration of this operation.

Note that a single tolerance $\delta$ is used to determine motion.
The theoretical value for $\delta$ is given by \eref{qdef} below.  We will
verify later that this value of 
$\delta$ is sufficiently small so  that two important properties hold:
\begin{description}
\item[Property 1 of $\delta$:] If a vertex $v'$ is the target
of a collapse-node move, then it should be uniquely determined, i.e., there
should not be two vertices $v'$ and $v''$ on the staying side of $b(v)$ that
are both within distance $\delta$ of $v$.
\item[Property 2 of $\delta$:] No two vertices on the moving side of $\big(e)$
for any $e$ should be collapsed to the same node on the staying side.
\end{description}
In a future extension of PINW to handle graded meshes, presumably the value
of $\delta$ would not be a single global value.

The affine transformations described above have the property that all of the
segments illustrated in Figure~\ref{fig:genPin} remain straight (collinear)
segments after the transformation.  
It is apparent that each collapse-node operation could cause many nodes
to move.  We will say that the one node $v$ that is displaced to match $v'$
is {\em directly} displaced.  The nodes moved by virtue of an affine
transformation induced by moving $v$ are said to be {\em indirectly} displaced.

A collapse-node operation, once executed, cannot be undone by future
collapse-node operations.  The reason
is that $v$ is never moved again.  It is never moved again directly since
it can be moved directly only when the tile 
$T=\bigtriangleup ABC$ that created it is processed.
It can also never be moved again indirectly since there is no tile in
lower levels of the hierarchy that contains
it except as a corner vertex, and corner vertices of a triangle $T'$ are not moved
when $T'$ is processed.  Similarly, $v'$ can never be moved again.  The reason
is that $v'$ is never moved directly (since it is on the staying side of its
big edge).  Any transformation that might move
$v'$ indirectly takes place at a level of the hierarchy higher than the level
of $T$.

We carry out all available collapse-node operations for all triangles
in the order described.  
Once all collapse-node operations are complete, we are left with the
collection of distorted leaf tiles, each of which may have one or more
hanging nodes.  These hanging nodes are collinear with the endpoints
of the edges on which they hang
because, as noted above, we do not disturb any collinearity relationships
with collapse-node operations.  The hanging nodes are all at least $\delta$
apart from the corners and from each other.

For each of these distorted tiles, we compute its
Delaunay triangulation (including the hanging nodes).  The collection
of all of these Delaunay triangles forms a simplicial mesh 
that is the final output of PINW.

The running time of PINW is analyzed as follows.  Let
$n$ be the number of leaf tiles.  Then the total number of tiles
is $O(n)$, as is the total number of vertices and edges.  The heap
insertions and deletions require $O(n\log n)$ total operations.  Sorting all the lists
associated with big edges requires $O(n\log n)$ operations.  Looking
up a vertex in a sorted list requires $O(\log n)$ operations for binary
search, hence all of the lookups to see if a node should be collapsed
require $O(n\log n)$ operations.  

The recursive application
of affine transformations requires $O(nd)$ operations since
each vertex is transformed at most $d$ times, where
$d$ is the maximum depth of the forest associated with $\T^*$.
We claim $d=O(\log n)$.  It follows from Lemmas~\ref{minalt1} 
and~\ref{minalt2} in the next section that the minimum altitude
of a triangle at depth $k$ lies
between $\alpha_0 C^k$ and $\alpha_1 D^k$, where $\alpha_0,\alpha_1$ 
are lower and upper bounds on the minimum altitudes among root tiles, 
$D$ is an absolute
constant and $C$ is a scalar depending on the worst aspect ratio among
root tiles.  This means that a leaf tile can be at most
a factor $\log D/\log C$ (asymptotically)
deeper in the forest than any other leaf tile.  Thus, all leaves
have depth $O(\log n)$.

Finally, the Delaunay triangulation operations in the last step
of the algorithm also
require $O(n\log n)$ operations total.
Overall, we see that PINW requires
$O(n\log n)$ operations.

\section{Analysis of aspect ratio}
\label{aspeffect}

In this section we analyze the aspect ratios of triangles produced
by PINW, showing that they are bounded above by a number that
depends only on the sharpest angle in the original polygon $\Omega$.
Before this analysis, we first
explain how to select the parameter $\delta$
described in the last section.  The parameter $\delta$ depends on
the minimum altitude of leaf tiles as will be
apparent from the theory developed here.
Let $\minalt(T)$ denote the minimum altitude of
triangle $T$.

\papselect{
\begin{lemma}
Let $T$ be a triangle and let $a=\minalt(T)$.  Then $T$
can be enclosed between two parallel lines at distance $a$ apart.  Conversely,
if $T$ can be enclosed between two parallel lines at distance $a$ apart,
then $\minalt(T)\le a$.
\label{parlem}
\end{lemma}
\begin{proof}
The first part of the lemma is quite trivial: draw a line through the
longest side length of $T$ and a parallel line through the opposite
vertex.  These lines are distance $a$ apart.  The argument for the converse is
as follows.  Without loss of generality, let the two lines be
parallel to the $x$-axis.  Let the vertices of $T$ be numbered
$v_1,v_2,v_3$ such that $v_1$ is closest to the bottom line (i.e., has
minimal $y$-coordinate among the three vertices) and $v_2$ to the top
line.  By reflecting if necessary, assume also that the $x$-coordinate
of $v_1$ is less than or equal to the $x$-coordinate of $v_2$.  Now draw the
line $v_1v_2$, which is a transverse to the two parallel lines.  If
$v_3$ is below (to the right) this line, then it is easy to see that
the entire triangle $T$ may be rotated clockwise about $v_1$ until
$v_1v_3$ becomes horizontal, and during this whole rotation, all three
vertices remain between the lines.  On the other hand, if $v_3$ is
above (to the left) of the line, then rotate $T$ clockwise
about $v_2$ until $v_2v_3$ becomes horizontal.

Once $T$ has been reoriented so that one of its edges is horizontal, the
claim is trivial since the altitude to the horizontal edge is a vertical
line segment and hence must have length
no more than $a$.
\end{proof}
}
{
}

\papselect{
\begin{corollary}
If $T_1, T_2$ are two triangles such that $T_1\subset T_2$,
then $\minalt(T_1)\le \minalt(T_2)$.
\label{contcor}
\end{corollary}
\begin{proof}
Draw the two parallel lines for $T_2$ as in the previous theorem;
clearly $T_1$ also lies between them.
\end{proof}
}
{
}

\begin{lemma}
Let $T$ be a triangle with vertices $v_1,v_2,v_3$.  Let $T'$ be the
triangle with vertices $v_1',v_2,v_3$.  Let $A$ be the 
unique affine transformation
that carries $T$ to $T'$.  Let $l$ be an arbitrary line segment.
Then
\begin{equation}
1-d/a\le \frac{\length(A(l))}{\length(l)} \le 1+d/a
\end{equation}
where $d=dist(v_1,v_1')$ and $a$ is the altitude of $v_1$
with respect to $v_2v_3$.
\label{stretchlem}
\end{lemma}

\papselect{
\begin{proof}
Without loss of generality, assume $a$ is 1 and $d$ is
replaced by $p=d/a$.
Furthermore, without loss of generality, let $T$ be positioned so that its
$v_2v_3$ edge is a subsegment of the $x$-axis and $v_1$ lies on the $y$-
axis (hence $v_1=(0,1)$ by the previous assumption).  
With these
assumptions, the affine tranformation $A$ in the question becomes a linear
transformation (i.e., no additive term) since the $x$-axis (and the origin
in particular) is invariant.  The transformation maps
maps $(1,0)^T$ to $(1,0)^T$ and $(0,1)^T=v_1$ to $v_1'$.  Let $r=v_1'-v_1$,
and let $r$ be written as $(\alpha,\beta)^T$ so that $\alpha^2+\beta^2=p^2$.
Then $A$ corresponds to the matrix
$$A=\left(
\begin{array}{cc}
1 & \alpha \\
0 & 1+\beta 
\end{array}
\right).
$$
The minimum and maximum distortion of a line segment under a linear
transformation is governed by the minimum and maximum singular values
of the transformation.
Thus, the question now hinges on the two singular
values of $A$. Notice that $A$ may be regarded
as a perturbation of the identity matrix, which has two singular
values equal to 1.
Therefore, by Corollary 8.6.2 of \cite{GVL},
the largest singular value of $A$ is at most $1+p$, and the smallest
singular value is at least $1-p$.  These values are attainable by taking
$\alpha=0$ and $\beta=\pm p$.
\end{proof}
}
{
This result follows in a fairly straightforward manner from 
a well-known theorem about how the singular values of a matrix
change under perturbation.
}

\begin{lemma}
Consider the generalized pinwheel subdivision illustrated
in Figure~$\ref{fig:genPin}$ of a triangle $T$.
Assume that $b\ge \min(.4,a)$
and $c-a\ge\min(.4,a)$.  Then
letting $T'$ be any one of the five subtriangles, we
have $\minalt(T')\le 0.9725\minalt(T)$.
\label{minalt1}
\end{lemma}

\noindent{\bf Remark 1.}
The assumptions are valid for all tiles produced by PINW.
For root tiles, we have ordered the angles $a\le b\le c$,
and we know $c-a\ge .4$ because of the preliminary step of splitting
near-equilateral triangles.
For conjugates of root triangles, say $a=a'$, $b=c'-a'$, and $c=\pi-c'$
where $a',b',c'$ are the angles of a root tile, we know also
$c-a=\pi-c'-a'=b'\ge a$ and that $b=c'-a'\ge .4$.

\noindent{\bf Remark 2.}
The factor $0.9725$ is due to our proof technique and appears to be
an overestimate.  A search over a fairly dense grid of 
possible angles satisfying
the hypotheses of the theorem 
indicates that the true bound is closer to $0.918$.

\papselect{
\begin{proof}
We start by observing that $|BC|/|AB|=\sin a/\sin c$ by the
law of sines.  We know that either $c\ge 2a$ or $c\ge a+.4$.
Furthermore, we know that either $\pi-c\ge 2a$ or $\pi-c\ge a+.4$
since $\pi-c=a+b$.  Now consider two cases.  Case 1 is that
$c\ge \pi/2$.  Define $\bar{c}=\pi-c$, so that $\sin \bar c=\sin c$,
$\bar c\ge\min(2a,a+.4)$ and $\bar c\le \pi/2$.  On the interval
$[0,\pi/2]$, the sine is increasing.  In
the subcase that $\bar c\ge 2a$, we have $\sin \bar c\ge \sin 2a=2\sin a\cos a
\ge \sqrt 2\sin a$.  The last inequality follows because $a\le \pi/4$
(by assumptions that $\bar c\le \pi/2$ and $\bar c\ge 2a$) so 
$\cos a\ge 1/\sqrt{2}$.  The other subcase is that $\bar c\ge a+.4$.
Since sine is concave and increasing on $[0,\pi/2]$
the worst case (maximum value) for $\sin a/\sin \bar c$
is when $a=\pi/2-.4$ and $\bar c=\pi/2$, so $\sin a /\sin \bar c\le .922$.
Thus, $|BC|/|AB|\le .922$.  The other case is $c\le \pi/2$.  This case
is handled by the same argument, except using $c$ in place of $\bar c$.

Observe that subtriangle V, which is denoted $T_{\rm V}$, is similar to
$T$ except scaled by a factor $|BC|/|AB|$.  This proves
$\minalt(T_{\rm V})\le .922\minalt(T)$.

Next, by similarity, $|BF|/|BC|=|BC|/|AB|$ hence $|BF|/|AB|\le .849$.
This means $|AF|/|AB|\ge .151$.  Since $\bigtriangleup ADF$ is isosceles,
$|AD|/|AF|\ge .5$ and $|DF|/|AF|\ge .5$ hence
$|AD|/|AB|\ge .075$ and $|DF|/|AB|\ge .075$.
Next, by the law of sines applied to $\bigtriangleup CDF$, 
$|DC|/|DF|=\sin b/\sin(c-a)$.  We now take three cases:
either $b<.4$, $b\in[.4,\pi/2]$, or $b>\pi/2$.
In the first case $b\ge a$ since the assumption in the lemma is $b\ge\min(.4,a)$.
Also, $c-a\ge \pi/2$ since $c-a+2a+b=\pi$ and $2a+b<1.2$ by the
assumption for this case, so $c-a\ge\pi-1.2$.  This means
$\sin(c-a)=\sin(\pi-(c-a))=\sin(2a+b)$, with $2a+b\le \pi/2$.  Next,
$2a+b\le 3b$ so $\sin(c-a)=\sin(2a+b)\le\sin(3b)$ (since $2a+b\le 3b$
and sine is increasing on $[0,\pi/2]$), so $\sin b/\sin(c-a)\ge\sin
b/\sin(3b)$.  Now $\sin(3b)=\sin b(3\cos^2 b-\sin^2 b)$.  Thus, $\sin
b/\sin(c-a)\ge 1/(3\cos^2 b-\sin^2b)$. Since $\sin^2$ is increasing
while $\cos^2$ is decreasing on $[0,.4]$, the minimum value of this
fraction is when $b=0$, so $\sin b/\sin(c-a)\ge 1/3.$ In the second
case, $b\in[.4,\pi/2]$.  This means $\sin b\in[\sin(.4),1]$ and in
particular, $\sin b\ge .389$ so $\sin b/\sin(c-a)\ge .389$.  The last
case is $b>\pi/2$, which implies $\sin(b)=\sin(\pi-b)=\sin(c+a)$.  So
the quantity to analyze is $\sin(c+a)/\sin(c-a)$.  Since the angles in
the numerator and denominator are both less than $\pi/2$ (because
$b>\pi/2$) and $c+a>c-a$, we conclude that $\sin b/\sin(c-a)\ge 1$.

Thus, in all cases, we conclude that $|DC|/|DF|\ge 1/3$.  This means that $|DC|/|AB| \ge .025$.

Next, observe that $|EF|/|DF|=\sin(c-a)/\sin c$ by the law of sines
applied to $\bigtriangleup DEF$.  Again, we take three cases.  If
$c-a<.4$ (and hence $c-a\ge a$ i.e., $c\ge 2a$, i.e., $a\le c/2$),
then $2a-a< .4$ i.e, $a\le .4$ so $c\le a+.4\le .8$.  Since all these
angles are in $[0,\pi/2]$, $\sin(c-a)\ge \sin(c-c/2)=\sin(c/2)$ and
$\sin(c/2)/\sin(c)\ge 1/2$ in this range by the convexity of $\sin$.
The second case is $c-a\in[.4,\pi/2]$.  In this case, $\sin(c-a)\ge
.389$ so $\sin(c-a)/\sin c \ge .389$.  The last case is $c-a\ge\pi/2$.
This implies that $\sin(c-a)=\sin(\pi-c+a)=\sin(2a+b)$.  The
denominator becomes $\sin(c)=\sin(\pi-c)=\sin(a+b)$.  So we are
analyzing $\sin(2a+b)/\sin(a+b)$, which exceeds 1 since all the angles
in question are in $[0,\pi/2]$.  Thus, in all cases, $|EF|/|DF| \ge
.389$ so $|EF|/|AB|\ge .0292$.

Now we have enough inequalities to analyze $\minalt(T_{\rm I})$ where
$T_{\rm I}$ denotes $\bigtriangleup ADE$.  Observe that this triangle is similar to
$T$.  Its longest side is either $AE$ or $AD$ (but not $DE$, since
$a<c$).  If its longest side is $AD$, then we see that
$|AD|=|AC|-|DC|\le |AC|-.025|AB|\le |AB|-.025|AB|=.9725|AB|$.  (Here
we used the fact that $|AB|\ge |AC|$, which follows from the
hypothesis of this case that $|AD|\ge|AE|$ plus the similarity of
$\bigtriangleup ADE$ to $\bigtriangleup ABC$.)  Thus, $\bigtriangleup ADE$ 
is similar to $\bigtriangleup ABC$ but is a factor
$.9725$ or less scaled down.

The other case is when the longest side of $ADE$ is $AE$.  In this case,
$|AE|=|AB|-|EB|\le |AB|-|EF|\le (1-.0292)|AB|\le (1-.0292)|AC|$.  
The inequality
$|AB|\le |AC|$ follows from the assumption that $|AD|\le |AE|$ and similarity.
Thus, $ADE$ is similar to $ABC$ but is scaled down by factor less than
$.9708$.  This concludes the analysis of  $\minalt(T_{\rm I})$.
This same analysis applies to $T_{\rm III}$, since $T_{\rm III}$ is
congruent to $T_{\rm I}$ (because $\bigtriangleup ADF$ is isosceles).

Next, observe that $|AE|/|AD|=\sin b/\sin c$ by the law of sines
applied to $\bigtriangleup ADE$.  Again, we take three cases.  If
$b<.4$ (and hence $b\ge a$), then $c>\pi/2$ as above so
$\sin c=\sin(\pi-c)=\sin(a+b)$, so the quantity to analyze is $\sin
b/\sin(a+b)$.  Using analysis like before, including steps like
$\sin(a+b)\le \sin(2b)=2\sin b\cos b\le 2\sin b$, we conclude that
$\sin b/\sin c\ge 1/2$ in this case.  If $b\in[.4,\pi/2]$, then we
conclude again that $\sin b/\sin c\ge .389$.  Finally, if $b>\pi/2$,
then $\sin b = \sin(\pi-b)=\sin(c+a)$, so again $\sin b/\sin c>1$.
Thus, in all cases, $|AE|/|AD|\ge .389$.  Since $|AD|/|AB| \ge .075$,
we conclude that $|AE|/|AB|\ge .029$.

Next, we analyze $T_{\rm II}$, that is, $\bigtriangleup DEF$.  Observe
that this triangle is similar $\bigtriangleup CFA$, and 
$\minalt(\bigtriangleup CFA)\le
\minalt(\bigtriangleup ABC)$ by Corollary~\ref{contcor}.  The corresponding side to
$AF$ is $FE$.  We have $|FE|=|AF|-|AE|\le |AF|-.029|AB| \le
|AF|-.029|AF|$.  Thus, $\minalt(T_{\rm II})\le 0.971\minalt(T)$.

Finally, we analyze $T_{\rm IV}$, which is $\bigtriangleup CGD$.  This
triangle is also similar to $\bigtriangleup CFA$.  
We showed earlier that $|CD|\le
.963|AC|$, and $|CD|/|AC|$ is the ratio of similarity between these
triangles.
\end{proof}
}
{
\begin{proof}
Due to space limitations, we merely sketch out the main ideas
of the proof.  Using the law of sines and the assumptions, we
prove many inequalities about the side lengths in
Figure~\ref{fig:genPin} such as $|BC|\le .922|AB|$.
With enough inequalities like this, we can then argue using
principles of similar triangles about the minimum altitude.
\end{proof}
}

The following lemma is like the previous one except with an
inequality in the opposite direction.
\begin{lemma}
Consider the generalized pinwheel subdivision illustrated
in Figure~\ref{fig:genPin} of a triangle $T$.
Assume that $b\ge \min(.4,a)$
and $c-a\ge\min(.4,a)$.  Then
letting $T'$ be any one of the five subtriangles, we
have $\minalt(T')\ge p\minalt(T)$, where for subtriangles I, II, III, IV,
$p\ge 0.0044$ and for subtriangle V, $p\ge \sin a$.
\label{minalt2}
\end{lemma}

\noindent{\bf Remark.}  
The factor $0.0044$ is due to our proof technique and appears to be
an underestimate.  A search over a fairly dense grid of 
possible angles satisfying
the hypotheses of the theorem 
indicates that the true bound is closer to $0.125$.

\papselect{
\begin{proof}
Again, we consider the five subtriangles and reuse some of the inequalities
in the preceding proof.  Starting with $T_{\rm I}$, which is similar to $\bigtriangleup ABC$,
recall that 
$|AD|\ge .075|AB|$, hence by similarity, $\minalt(\bigtriangleup ADE)\ge .075\minalt(\bigtriangleup ABC)$.
The same bound applies to $T_{\rm III}$, which is congruent to $T_{\rm I}$.

For $T_{\rm V}$, by the law of sines $|BC|/|AB|= \sin a/\sin c\ge \sin a$.
Thus, by similarity, $\minalt(T_{\rm V}) \ge \minalt(T)\cdot \sin a$.

Next, consider triangle $\bigtriangleup ACF$.  
In the previous proof we showed that $|AF|/|AB|\ge .151$, which means that 
if $\bigtriangleup ACF$ were
dilated by $1/.151$, it would completely cover $\bigtriangleup ABC$.  
Therefore, 
by Corollary~\ref{contcor},
$\minalt(\bigtriangleup ACF)\ge .151\minalt(\bigtriangleup ABC)$.  
Next, we showed that $|EF|/|AB|\ge .0292$ so $|EF|/|AF|\ge .0292$.  Since $T_{\rm II}$ is
similar to $\bigtriangleup ACF$, we conclude that $\minalt(T_{\rm II})\ge .0292
\minalt(\bigtriangleup ACF)
\ge .0292\cdot .151\minalt(\bigtriangleup ABC)$.  Note that $.0292\cdot 0.151\ge 0.0044$.

Finally, to analyze $T_{\rm IV}$, we need to develop new inequalities.
Recall we have already shown that $|DF|/|AB|\ge .075$.  Since
$\bigtriangleup DGF$ is similar to $\bigtriangleup ACB$, 
this implies $|DG|\ge .075|AC|$.  Meanwhile, we know $|AC|\ge 0.5|AF|$
since $|AC|$ is not the shortest side of $\bigtriangleup ACF$ (because
it is opposite an angle of size $a+b$, which is greater than the angle
at $A$ of size $a$).  Thus, $|DG|\ge .037 |AF|$.  This means by similarity
of $T_{\rm IV}$ to $\bigtriangleup ACF$ that $\minalt(T_{\rm IV})\ge .037\minalt(\bigtriangleup ACF)
\ge .037\cdot .151\minalt(\bigtriangleup ABC)$.  
\end{proof}
}
{
Due to space limitations, the proof is omitted.
}

\begin{lemma}
Let $s$, $t$ be positive numbers such that $s<1$ and $t<1$, and 
let $k$ a positive
integer.  Then
$$\prod_{i=0}^\infty (1-st^i)^k \ge 1-ks/(1-t)$$
and
$$\prod_{i=0}^\infty (1+st^i)^k \le \exp(ks/(1-t)).$$
\label{prodlem}
\end{lemma}
\papselect{
\begin{proof}
The first inequality follows because $(1-a)(1-b)\ge 1-a-b$ for $a,b\in[0,1]$,
Applying this repeatedly, $k$ times for each factor
in the product, $(1-s)^k(1-st)^k\cdots(1-st^n)^k\ge 1-ks-kst-\cdots-kst^n\ge
1-ks\sum_{i=0}^\infty t^i=1-ks/(1-t)$.

The second inequality follows by taking logs and using the inequality
$\log(1+x)\le x$:
\begin{eqnarray*}
\log ((1+s)^k\cdots(1+st^n)^k) & =&k\log(1+s)+k\log(1+st)+\cdots+k\log(1+st^n) \\
&\le &ks+kst+\cdots+kst^n \\
&\le & ks\sum_{i=0}^\infty t^i\\
&=&ks/(1-t).
\end{eqnarray*}
\end{proof}
}
{
}

We now explain how to choose $\delta$ for the main algorithm.  We
set it to be
\begin{equation}
\delta=\frac{\min\{\minalt(T): T\in {\mathcal T}_*\}}{1460}.
\label{qdef}
\end{equation}
The minimum altitudes in this definition are measured before any collapse-node
operations begin.
This choice of $\delta$ makes all the theorems
work but leads to poorer aspect ratio (by a constant factor) than seems
necessary.  So instead, PINW 1.0 chooses $\delta$ dynamically based
on the singular values of the affine transformations that
could be applied during collapse-node operations.
This heuristic seems
to work well in practice.

The following theorem bounds the effect of all collapse-node operations,
both direct and indirect.

\begin{theorem}
Let $T$ be a tile in the hierarchy generated
by PINW, and let $A$ be the composition of all the affine tranformations
applied directly to vertices of $T$ and indirectly to those
vertices via ancestors in the hierarchy.  
Let $\alpha=\minalt(T)$ (prior
to any node movement).
Let $l$ be a line segment lying in $T$.  Assume $\delta$ is chosen according to \eref{qdef}.
Then $\length(A(l))/\length(l)$ lies between
$$
\left(1-\frac{\delta}{0.75\alpha}\right)^3\cdot
\left(1-\frac{0.9725\delta}{0.75\alpha}\right)^3\cdot
\left(1-\frac{0.9725^2\delta}{0.75\alpha}\right)^3\cdots 
$$ 
and
$$
\left(1+\frac{\delta}{0.75\alpha}\right)^3\cdot
\left(1+\frac{0.9725\delta}{0.75\alpha}\right)^3\cdot
\left(1+\frac{0.9725^2\delta}{0.75\alpha}\right)^3\cdots
$$
\label{affstretch}
\end{theorem}
\papselect{
\begin{proof}
The proof of this theorem is by induction.  The induction base is that for
a root tile, there are no collapse-node operations so $A(l)=l$.
For a nonroot tile $T$, let $T'$ be its parent triangle.

By the induction hypothesis,
the total distortion of a segment in $T'$ prior to the processing of
the vertices created within $T'$ is between
$$
\left(1-\frac{\delta}{0.75\alpha'}\right)^3\cdot
\left(1-\frac{0.9725\delta}{0.75\alpha'}\right)^3\cdot
\left(1-\frac{0.9725^2\delta}{0.75\alpha'}\right)^3\cdots
$$
and
$$
\left(1+\frac{\delta}{0.75\alpha'}\right)^3\cdot
\left(1+\frac{0.9725\delta}{0.75\alpha'}\right)^3\cdot
\left(1+\frac{0.9725^2\delta}{0.75\alpha'}\right)^3\cdots,
$$
where $\alpha'=\minalt(T')$ (with minalt measured prior to any node movement).
Referring to Figure~\ref{fig:genPin} and regarding $T'=ABC$ and $T$ as one of
I, II, III, IV or V,
we consider next the direct displacements of $D$, $E$, $F$ due to
collapse-node operations.

By Lemma~\ref{minalt1}, $0.9725\alpha'\ge \alpha$.
Thus, for tile $T$, the distortion prior to the three direct
displacements of $D,E,F$ is bounded between
\begin{equation}
\left(1-\frac{0.9725\delta}{0.75\alpha}\right)^3\cdot
\left(1-\frac{0.9725^2\delta}{0.75\alpha}\right)^3\cdot
\left(1-\frac{0.9725^3\delta}{0.75\alpha}\right)^3\cdots
\label{str1}
\end{equation}
and
\begin{equation}
\left(1+\frac{0.9725\delta}{0.75\alpha}\right)^3\cdot
\left(1+\frac{0.9725^2\delta}{0.75\alpha}\right)^3\cdot
\left(1+\frac{0.9725^3\delta}{0.75\alpha}\right)^3\cdots.
\label{str2}
\end{equation}
By Lemma~\ref{prodlem} with $k=3$, $s=\delta/(0.75\alpha)$ and
$t=0.9725$, infinite product \eref{str1} is greater than
or equal to $1-3\delta/(0.75\alpha(1-0.9725))$ which simplifies
to $1-146\delta/\alpha$.  If we assume 
that $\delta$ satisfies \eref{qdef}, then
this quantity is greater than $0.9$.  We now apply the
three collapse operations of $D,E,F$ to $T$, say in this order.  (Not all three necessarily
affect $T$; for example, if $T$ is I in the figure, then moving $F$ does
not affect $T$.)
To compute the distortion of $l$ requires knowledge of the minimum
altitude of $T$ at the point of the algorithm
when the collapse-node operation is applied. 
However, because the distortion so far
is greater than $0.9$, we know that the altitude
at this step is at least $0.9\alpha$ for
movement of $D$, which is of size $\delta$.  Therefore, the  movement of $D$
applies a new distortion between $1-\delta/(0.9\alpha)$
and $1+\delta/(0.9\alpha)$
by Lemma~\ref{stretchlem}. 
By \eref{qdef}, this
quantity is bounded between $0.95$ and $1.05$.
Therefore, the minimum altitude of $T$ when the 
collapse-node operation for $E$ is applied is at least $0.9\alpha \cdot 0.95=0.855\alpha$.
Thus, the collapse-node operation on $E$ applies another distortion
between $1-\delta/(0.855\alpha)$ and $1+\delta/(0.855\alpha)$.
Again, by \eref{qdef}, this quantity is bounded between
$0.95$ and $1.05$.  So after the collapse-node operation on $F$,
the minimum altitude of $T$ is at least $0.855\alpha\cdot 0.95=0.812\alpha$.
Combining these three distortions with the distortions from higher-level
collapse-node operations given by \eref{str1} and \eref{str2} shows
that the minimum and maximum distortion of
a segment after the collapse-node operations
involving $T$ and its ancestors lies between
\begin{eqnarray*}
\lefteqn{(1-\delta/(0.9\alpha))(1-\delta/(0.855\alpha))(1-\delta/(0.812\alpha))} \\
& &\cdot
\left(1-\frac{0.9725\delta}{0.75\alpha}\right)^3\cdot
\left(1-\frac{0.9725^2\delta}{0.75\alpha}\right)^3\cdot
\left(1-\frac{0.9725^3\delta}{0.75\alpha}\right)^3\cdots
\end{eqnarray*}
and
\begin{eqnarray*}
\lefteqn{(1+\delta/(0.9\alpha))(1+\delta/(0.855\alpha))(1+\delta/(.812\alpha))}
\\
& &
\cdot
\left(1+\frac{0.9725\delta}{0.75\alpha}\right)^3\cdot
\left(1+\frac{0.9725^2\delta}{0.75\alpha}\right)^3\cdot
\left(1+\frac{0.9725^3\delta}{0.75\alpha}\right)^3\cdots.
\end{eqnarray*}
We can underestimate the first factor and overestimate the second by replacing
$0.9$, $0.855$ and $0.812$ all with $0.75$.  This proves the theorem.
\end{proof}
}
{
Due to space limitations, we defer the proof of this theorem to
the full paper.  
}

We now consider Properties 1 and 2
in
Section~\ref{algo}.  
Since the minimum altitude of a triangle is less than or equal to
its shortest side length, and since the minimum altitude of any tile decreases by
at most $0.75$, the previous result shows that $\delta$ is sufficiently small
so that no two nodes can be collapsed to the same node, and no node can have
more than one choice of where it should be collapsed.

Furthermore, when we are finished with collapse-node operations, all hanging nodes
are at least $\delta$ apart and at least $\delta$ from corners.  Again, this is because
the shortest side length is bounded below by the smallest altitude, and the smallest
altitude is bounded below by a large constant multiple of $\delta$.

We now consider the aspect ratio of the triangles in the mesh produced by PINW.
We define the {\em aspect ratio} of a triangle to be
the square of the longest side length of the triangle divided by
its area.  Since the area is half the product of the longest side length
and the minimum altitude, an equivalent definition is
twice the longest side length over the minimum altitude.

\papselect{
The following lemma gives another characterization of aspect ratio equivalent
up to a constant factor
as well as a useful property of aspect ratios.

\begin{lemma}
Let $T$ be a triangle with aspect ratio $a$.  

(a) Let $\theta$ be
the minimum angle of $T$.  Then there exists two universal constants
$c_1,c_2$ such that $c_1 a \le 1/\theta \le c_2 a$.

(b) Let $l_1,l_2$ be two distinct
side lengths of $T$.  Then $l_1/l_2 \le a/2$.
\label{asplem}
\end{lemma}

\begin{proof}
Because this lemma is well-known (see, e.g., \cite{Knupp}),
we omit the full proof.
For (a), let $H$ be the length of the longest edge of $T$.
The proof of (a) follows by noting that there exists a right triangle 
one of whose legs has length $H$ and one of whose angles is $\theta$ that
contains $T$.  On the other hand, the same right triangle contracted
by a factor of $1/2$
is contained in $T$.  For (b), we observe
that $l_1l_2\ge 2\area(T)$, i.e., 
$l_2\ge 2\area(T)/l_1= 2l_1(\area(T)/l_1^2)\ge 2l_1/a$.
\end{proof}
}
{
}

The first step of PINW, which performs a
preliminary triangulation of $\Omega$ using Triangle,
outputs triangles that have their aspect
ratios bounded above. 
The reason is that Triangle is a guaranteed-quality mesh
generation algorithm that will put sharp angles into its output only
when the input polygon has very sharp angles. 
Thus, the small angles of all the initial triangles have a
lower bound.  (The reciprocal of the smallest angle of a triangle
is within a constant factor of the aspect ratio definition given
in the previous paragraph.)
The operation of subdividing
at in-centers done to obtain ${\mathcal T}_0$ 
from Triangle's output does not increase the longest side length,
and reduces the area by at most a constant factor.  Hence the triangles
in ${\mathcal T}_0$ still have bounded aspect ratio.

Next, we consider the tiles in ${\mathcal T}_f$, that is, the leaf
tiles.  Each of these is similar to a root tile or its conjugate.
In a preliminary step, we ensured that $c-a$ is bounded below
for all conjugates of root tiles.  Therefore, the leaf tiles all have bounded
aspect ratio.

In more detail, the smallest angle of each conjugate leaf tile is
either $a$, where $a$ is the smallest angle of a root tile, or is
$c-a$, where $a$ is the smallest and $c$ is the largest angle of
a root tile.  But we have ensured that $c-a>.4$ by our preliminary
splitting rule.  Thus, if the smallest angle of a conjugate tile is
$c-a$, this means that the conjugate tile has a universal upper
bound on its aspect ratio.

Now, we consider the effect of collapse-node operations.  
\begin{lemma}
After all collapse-node operations are complete, the aspect ratio
of any leaf tile has increased (compared to its value
prior to all collapse-node operations)  by at most a factor of $1.22$.
\end{lemma}
\papselect{
\begin{proof}
As explained in the proof of Theorem~\ref{affstretch}, 
the smallest distortion due to all collapse-node operations
for any segment in any
leaf tile is 0.90 or greater. Pick a tile $T$ and
let $\alpha$ be the initial altitude.
Applying the second part of Lemma~\ref{prodlem}
to the bound in the theorem shows that
the maximum distortion of $T$ is $\exp(3\delta/(0.9\alpha(1-0.9725)))$,
which by \eref{qdef} is at most $1.09$.  
Since the aspect ratio is the twice the longest side length divided by the minimum
altitude, and the longest side went up by at most 1.09 while the minimum
altitude changed by a factor at least 0.90, the new aspect ratio
is bounded by $1.22$ times the old.
\end{proof}
}
{
}

For this theorem and the remainder of the section, let $R_1$ denote
the maximum aspect ratio among root tiles and their conjugates.  As noted
above, because of the properties of Triangle, $R_1$ is bounded above by
a constant multiple of the reciprocal of the sharpest angle of $\Omega$.

\begin{theorem}
Assume that
no root tile is in ${\mathcal T}_f$
(i.e., each triangle in ${\mathcal T}_0$ is
split at least once by the PINW subdivision procedure).  Then,
prior to collapse-node operations,
the maximum value of the minimum altitude among 
all leaf tiles is no more than $cR_1$ times the minimum value
of the minimum altitude among all leaf tiles,
where $c$ is a universal
constant.
\label{maxminthm}
\end{theorem}
\begin{proof}
Recall that the tile selected for splitting at any given step is
the one with the maximum minimum altitude.  Thus, when the subdivision
procedure terminates, the tile at the top of the heap will be the
leaf tile with the maximum minimum altitude among all leaf tiles.  Say this tile
is $T$ and its minimum altitude is $\alpha$.
Now consider any other leaf tile $T'\in{\mathcal T}_f$.  
Because of the assumption that
no tile from ${\mathcal T}_0$ is a leaf tile, this tile $T'$ must have arisen
from a subdivision of some other tile $T''$.  Because of the heap order, the
minimum altitude  $\alpha''$ of $T''$ exceeds $\alpha$.
By Lemma~\ref{minalt2}, this means that the minimum altitude $\alpha'$ of $T'$
is at least $\min(0.0044,\sin a)\alpha''$.  Note that $\sin a \ge c/R_1$ since $a$ is
an angle of a root tile.   Thus, $\alpha'\ge (c/R_1)\alpha''\ge(c/R_1)\alpha$.
\end{proof}

We now come to the main result for this section about the aspect ratio of the
triangles generated by PINW.

\begin{theorem}
Each triangle in the simplicial mesh output by PINW has aspect ratio at most $cR_1^3$,
where $c$ is a universal constant and $R_1$ was defined above to be the 
largest aspect ratio
among root tiles.
\end{theorem}

\papselect{
\begin{proof}
Let $T$ be a leaf tile.  Let $\alpha$ be its minimum altitude and
$M$ its longest side length prior to
any collapse node operations.  As already observed in the proof of
Theorem~\ref{affstretch}, at the end of collapse-node operations, its
minimum altitude is at least $0.9\alpha$ and
its longest side length at most $1.09M$.  The edges of $T$ contain
hanging nodes.  The distance between any pair of adjacent hanging
nodes or between a hanging node and corner is at least $\delta$.  This
is because the shortest side length of any leaf tile is a sizable
constant multiple of $\delta$, so no leaf tile edge could ever shrink
below $\delta$.

Recall from \eref{qdef} that $\delta$ is a constant multiple of
$\alpha_{\rm min}$, the minimum altitude among all leaf tiles.  By
Theorem~\ref{maxminthm}, this implies $\delta\ge c\alpha/R_1$, where
$c$ is a universal constant and $\alpha=\minalt(T)$.

Now, let $\tau$ be a triangle output by the Delaunay triangulation of
$T$ and consider the sharpest angle of $\tau$.  Let $e=v_1v_2$
be edge of $\tau$ opposite the sharpest angle.  There are
two cases: either $v_1,v_2$ lie on the same edge of $T$ (i.e., they
are consecutive hanging nodes or a hanging node and a corner node),
or they are on different edges.

Start with first case.  Let $e=v_1v_2$ be
the edge of $\tau$ lying on an edge of $T$. 
For the rest of this case, let $T=\bigtriangleup ABC$ such that
$v_1v_2\subset BC$ and such that the order of these vertices is
$B,v_1,v_2,C$.
As mentioned above,
$|e|\ge c\alpha/R_1$.   Let $p$
be the vertex of $\tau$ opposite $e$.  By definition of the Delaunay triangulation,
$p$ is the first vertex hit by an expanding circle that contains both
endpoints of $e$.  This circle, if expanded further,
will eventually hit $A$, the vertex of $T$
opposite the edge of $T$ containing $e$.  
Either $\angle Av_1v_2$ is acute or $\angle Av_2v_1$ is acute; without loss of
generality, assume the former.
The angle $\angle v_1Av_2$
is at least $c|e|/(R_1 M)$ by the law of sines:
$\sin \angle v_1Av_2=\sin\angle Av_1v_2 |e|/|Av_2|
\ge \sin\angle Av_1v_2 |e|/(1.09M)$.  Meanwhile, 
$\sin\angle Av_1v_2$ is bounded below by $c/R_1$ since $\angle Av_1v_2$ is
bounded below by $\angle ABC$ but is less than $\pi/2$.  Thus,
$\sin\angle v_1Av_2\ge c|e|/(R_1M)$.  Next, $M\le \alpha R_1$
so $\sin\angle v_1Av_2\ge c|e|/(\alpha R_1^2)$.  Finally, 
$|e|/\alpha\ge c/R_1$ as noted in the previous paragraph.
Therefore, $\sin\angle v_1Av_2\ge c/R_1^3$.
This is the angle formed by $v_1Av_2$.  The actual angle of $\tau$
opposite $e$ is $v_1pv_2$.  But this angle is greater than or equal
to $v_1Av_2$, since the expanding Delaunay circle encounters $p$ before
it encounters $A$ (or else $p=A$).

Next, let us consider the case that $v_1$ and $v_2$, the endpoints
of the edge of $\tau$ opposite its sharpest angle, do not lie on the
same edge of $T$.
Let the three vertices of $T$ be $A,B,C$ and let $A$
be the $T$-vertex that is the common endpoint of the two
$T$ edges that contain $v_1$ and $v_2$ respectively, while we let
$B$ be the $T$-vertex such that $AB$ contains $v_1$ and
we let $C$ be the $T$-vertex such that $AC$ contains $v_2$.
Consider $\angle v_1Bv_2$.  Since $A$, $v_1$ and $B$ are collinear,
this angle is equal to $\angle ABv_2$.  By the law of sines
applied to $\bigtriangleup ABv_2$, we have
$\sin\angle ABv_2=|Av_2|\sin\angle A/|Bv_2|$.
Now we apply the following inequalities: $|Av_2|\ge \delta$,
$\sin\angle A\ge c/R_1$, and $|Bv_2|\le M$  to conclude
that $\sin\angle ABv_2=\sin\angle v_1Bv_2\ge c\delta/(R_1M$).
This was the same inequality derived in the previous case, and yields
the conclusion that $\sin\angle v_1Bv_2\ge c/R_1^3$.  Arguing
again as in the previous case, the actual Delaunay triangle containing
$v_1v_2$ may not have $B$ as its third vertex, but if it has any other vertex $w$,
then $\angle v_1wv_2$ is greater than $\angle v_1Bv_2$ by considering the
expanding circle property.
\end{proof}
}
{
Due to space limitations, we omit the proof.  In brief, the idea
is that we consider the
sharpest angle of a triangle $\tau$ in the Delaunay triangulation.
There are two cases: either the edge $e=v_1v_2$ opposite the sharp angle lies
on an edge $E$ of $T$ or it connects two edges of $T$.  Consider only
the first case.  We can bound the sharp angle of $\tau$ by the sharp
angle $v_1vv_2$ of the triangle $\tau'$ 
resulting from joining $e$ to $v$, where $v$ is
the vertex of $T$ opposite $E$.  We find that the aspect ratio of
$\tau'$ is proportional to $R_1^3$.  One factor of $R_1$ arises because
$|e|/\diam(T) \ge c/R_1$.  The other two factors of $R_1$
arise because if $R_1$ is large, $v$ could make a large angle (close
to $\pi$) with $e$.
}

\section{Isoperimetry of the final mesh}

We have already proved in Theorem~\ref{isoperim1}
that the tiling of a triangle by our generalized pinwheel
subdivision has the isoperimetric property.  It is
straightforward to extend this result to the collection
of all leaf tiles.

\begin{theorem}
Let ${\mathcal T}_0,{\mathcal T}_1,\ldots,$ be the sequence of tilings
of $\Omega$ generated by the PINW algorithm as follows.  For each
$n$, ${\mathcal T}_n$ is the set of leaf tiles of $\Omega$ generated
by PINW when the user-specified size requirement is $\delta_n$ such
that 
$\delta_n\rightarrow 0$ as $n\rightarrow \infty$.  Then for any 
distinct points $P,Q$ such that $P,Q\in\Sk({\mathcal T}_k)$ for some
$k$, 
we have 
$$\lim_{
\renewcommand\arraystretch{0.5}
\begin{array}{cc}
\scriptstyle n\rightarrow \infty\\
\scriptstyle n\ge k
\end{array}
}
\dist_{\Sk({\mathcal T}_n)}(P,Q)
=\Vert P-Q\Vert_\Omega.$$
\label{isoperim3}
\end{theorem}
\begin{proof}
This theorem follows from Theorem~\ref{isoperim2} and uses the same
proof technique.  Let $\Pi$ be the geodesic path from $P$ to $Q$ of
length $\Vert P-Q\Vert_\Omega$.  Since $\Omega$ is a polygon, 
$\Pi$ is composed of a finite number of line segments.
For each tile $T_i$ in 
${\mathcal T}_k$ that meets $\Pi$, consider the
small segment $P_iQ_i$ that is $T_i\cap \Pi$.  Then we use
Theorem~\ref{isoperim1} to argue that this small segment $P_iQ_i$
can be approximated arbitrarily accurately.
\end{proof}

This theorem can now be extended to the final
mesh output by PINW by analyzing
the effect of collapse-node operations on the isoperimetric
number.  (The Delaunay operations
do not disturb the isoperimetry result, since 
adding edges could only make the isoperimetric number decrease.)

The definition of isoperimetry implicit in Theorems~\ref{isoperim2} 
and~\ref{isoperim3} is not suitable for analyzing the output of PINW
because the meshes produced by PINW are not refinements of their
predecessors as the mesh size decreases.  This is because the
collapse-node operations move nodes differently depending on the
size of the leaf tiles.

Therefore, we use the
following definition.  An infinite sequence of simplicial 
meshes ${\mathcal M}_1,
{\mathcal M}_2,\ldots$ for a domain $\Omega$
has the {\em isoperimetric property} if for each
${\mathcal M}_i$ there is a subset $L_i$  of its vertices such that
the following two properties hold.  First, 
$L_i$ is asymptotically dense in $\Omega$ as $i\rightarrow\infty$,
i.e., for any $\epsilon>0$,
there is an $I$ such that for any $x\in \Omega$ and any $i>I$,
there exists a $v\in L_i$ such that $\Vert x-v\Vert\le \epsilon$.
Second,
$$\lim_{i\rightarrow \infty} \sup\left\{\frac{\dist_{\Sk({\mathcal M}_i)}(x,y)}{\Vert x-y\Vert_\Omega}: x,y\in L_i;x\ne y\right\} = 1.$$
\begin{theorem}
The family of meshes produced by the PINW algorithm has the isoperimetry
property described in the previous paragraph.
\label{isperim4}
\end{theorem}

\begin{proof}
To show that PINW has this property, take a sequence of $\epsilon_i$'s
tending to zero.  For each $i$, let ${\mathcal T}_i$ be a generalized
pinwheel subdivision of $\Omega$ such that each leaf cell has diameter less
than $\epsilon_i/2$.  Then let ${\mathcal T}_{i}'$ be a further subdivision
of ${\mathcal T}_i$ such that for any two 
distinct vertices of ${\mathcal T}_i$, $\dist_{\Sk({\mathcal T}_{i}')}(x-y)\le
(1+\epsilon_i/4)\Vert x-y\Vert_\Omega$.  The existence of such an
$\T_i'$ is established by Theorem~\ref{isoperim3}.
Let $\alpha'$ be the minimum altitude among leaf tiles
in ${\mathcal T}_i'$. 
Next, further refine ${\mathcal T}_i'$ to yield
a tiling ${\mathcal T}_i''$ with the property that
when $\delta$ is defined by \eref{qdef} for ${\mathcal T}_i''$,
(i.e., the ${\mathcal T}_*$ appearing in \eref{qdef} pertains to
${\mathcal T}_i''$), then this $\delta$ is sufficiently small
so that
\begin{equation}
\exp(3\delta/(0.75\alpha'(1-0.9725))) \le 1+\epsilon_i/4
\label{delrq1}
\end{equation}
and
\begin{equation}
1-\epsilon_i/4\le 1-3\delta/(0.75\alpha'(1-0.9725)).
\label{delrq2}
\end{equation}
Now finally, take ${\mathcal M}_i$ to be the simplicial mesh
output by PINW based on ${\mathcal T}_i''$, and take
$L_i$ to be the set of nodes of ${\mathcal M}_i$ that are displaced
copies of the nodes of ${\mathcal T}_i$.

First, we have to show that $L_i$ defined in this manner
is asymptotically dense.  The nodes of $L_i$ are the same as the
nodes of ${\mathcal T}_i$ after small displacements.  Since every
cell of ${\mathcal T}_i$ has diameter less than $\epsilon_i/2$,
this means that any point $x\in\Omega$
is distance at most $\epsilon_i/2$ from
a vertex of ${\mathcal T}_i$.  
The vertices of $L_i$ are slightly displaced, but no distance
$d$ decreases below $0.9d$ nor increases to more than
$1.09d$.  Therefore, for any $x\in \Omega$ the perturbed set $L_i$ contains
a point $v$ within distance $1.09\cdot\epsilon_i/2 < \epsilon_i$
of $x$.

Let $x,y$ be two distinct points in $L_i$.  The next task is to show that
$\dist_{\Sk({\mathcal M}_i)}(x,y)\le \Vert x-y\Vert_\Omega(1+\epsilon_i)$.
Let $x_0,y_0$ be the positions of $x,y$ in ${\mathcal T}_{i}''$ prior
to all distortions caused by collapse-node operations.  
Note that $x_0,y_0$ are vertices of ${\mathcal T}_{i}'$ and also of
${\mathcal T}_i$ by construction.  Therefore,
by construction of ${\mathcal T}_{i}'$, there is a path
$P_0$ in $\Sk({\mathcal T}_{i}')$ connecting $x_0$ and $y_0$
such that $\length(P_0)\le \Vert x_0-y_0 \Vert_\Omega\cdot (1+\epsilon_i/4)$.
Let the segments of $P_0$ be $e_1,e_2,\ldots,e_r$.
Let the image of $P_0$ after all the collapse-node operations 
with their attendant distortions are
applied be $P$, and the images of $e_1,\ldots,e_r$ be
$f_1,\ldots,f_r$.   Recall that the distortions that affect
a node $v$ of a tile $T$ in the hierarchy are those distortions
associated with $T$ and its ancestor tiles, but descendant tiles cannot
move $T$.  Therefore, by Theorem~\ref{affstretch}, all of the quantities 
$\Vert x-y\Vert / \Vert x_0-y_0\Vert_\Omega$, $\length(f_i)/\length(e_i)$, and 
$\length(P)/\length(P_0)$ lie between
$$
\left(1-\frac{\delta}{0.75\alpha'}\right)^3\cdot
\left(1-\frac{0.9725\delta}{0.75\alpha'}\right)^3\cdot
\left(1-\frac{0.9725^2\delta}{0.75\alpha'}\right)^3\cdots 
$$ 
and
$$
\left(1+\frac{\delta}{0.75\alpha'}\right)^3\cdot
\left(1+\frac{0.9725\delta}{0.75\alpha'}\right)^3\cdot
\left(1+\frac{0.9725^2\delta}{0.75\alpha'}\right)^3\cdots
$$
where the $\delta$ in this formula is given by \eref{qdef}
associated with ${\mathcal T}_{i}''$.  By Lemma~\ref{prodlem},
this interval is bracketed by
$$1-3\frac{\delta}{0.75\alpha'(1-0.9725)}$$
and
$$\exp(3\delta/(0.75\alpha'(1-0.9725))).$$
Then by \eref{delrq1} and \eref{delrq2}, this interval is
bracketed by $[1-\epsilon_i/4,1+\epsilon_i/4]$.
Thus,
\begin{eqnarray*}
\dist_{\Sk({\mathcal M}_i)}(x,y) &\le &
(1+\epsilon_i/4)\dist_{\Sk({\mathcal T}_i')}(x_0,y_0) \\
&\le &
(1+\epsilon_i/4)^2\Vert x_0-y_0\Vert_\Omega \\
&\le&
(1+\epsilon_i/4)^2\Vert x-y\Vert_\Omega/(1-\epsilon_i/4).
\end{eqnarray*}
Note that $(1+\epsilon_i/4)^2/(1-\epsilon_i/4)\le 1+\epsilon_i$
as long as $\epsilon_i\le 1/2$.  Thus, we have shown
that for all $x,y\in L_i$,
$\dist_{\Sk({\mathcal M}_i)}(x,y) \le (1+\epsilon_i)\Vert x-y\Vert_\Omega$.
\end{proof}

\section{Conclusions}
\label{conc}

\papselect{
We believe that this generalization of pinwheel tiling to meshing
polygonal regions would aid in modeling arbitrary crack
paths more accurately than the current meshing techniques.  Also, this
kind of substitutive mechanism for subdivision makes it easy for
adaptive meshing.  This work raises a number of interesting directions
for future research.  Among them are the following:
}
{
We believe that this generalization of pinwheel tiling to meshing
polygonal regions would aid in modeling arbitrary crack
paths more accurately than the current meshing techniques. 
This work raises a number of interesting directions for future research:
}

\begin{enumerate}
\item
The transformation of the tiling to the mesh had the effect of
increasing the aspect ratio significantly.
Is there a better way to carry out this transformation to
reduce the impact on aspect ratio?

\item
The convergence rate of the isoperimetric
number of the pinwheel tiling to 1, which was not analyzed
here, is known to be extremely slow even in the case of the 1:2
tiling.  Is there another approach to isoperimetry that converges faster?

\item
Consider a mesh generated by placing random points in the domain under
consideration and joining them with a Delaunay triangulation.  Is there
a limiting isoperimetric number for this family of meshes (with high 
probability)?

\item
Another way to construct a mesh of an arbitrary polygon
with limiting isoperimetric number equal to
1 is to use the 1:2 pinwheel subdivision for every coarse triangle
after subjecting it to a (potentially large) affine transformation.
This approach is simpler in certain respects than PINW.  For example,
the collapse-node operations for this algorithm need to be done
only at the boundaries of the coarse triangles.  The difficulty
with this approach is that it spoils the ``statistical rotational
invariance'' of the pinwheel tiling.  The statistical rotational
invariance property states that 
the set of possible
directions is covered at a uniform rate as subdivision proceeds.
We are unclear whether statistical rotational invariance is important
for cohesive interface modeling.  We suspect that our
construction of generalized pinwheels has statistical rotational
invariance but have no proof of this.

\item
Can any of this work be extended to three dimensions?
\end{enumerate}

\section{Acknowledgements}
We are grateful to Marshall Bern for originally telling us
about Radin's paper.  We also thank the two reviewers of the short
version of this paper submitted to the 2004 International
Meshing Roundtable for their helpful comments.

%\end{description}
\bibliography{pinbib}

\end{document}